\documentclass[lettersize,journal]{IEEEtran}
\usepackage{amsmath,amsfonts}
\usepackage{algorithmic}
\usepackage{algorithm}
\usepackage{array}
\usepackage[caption=false,font=normalsize,labelfont=sf,textfont=sf]{subfig}
\usepackage{textcomp}
\usepackage{stfloats}
\usepackage{url}
\usepackage{verbatim}
\usepackage{graphicx}
\usepackage{cite}
\usepackage{mathtools}
\usepackage{amssymb}
\usepackage{enumitem}
\usepackage{hyperref}
\usepackage{multirow}
\usepackage{float} 
\usepackage{wrapfig}
\usepackage{mathrsfs}

\begin{document}
	
	
	\title{A Deep Incremental Framework for Multi-Service Multi-Modal Devices in NextG AI-RAN Systems }
	
	\author{Mrityunjoy~Gain,~\IEEEmembership{Student Member,~IEEE,}
		Kitae~Kim,
		Avi Deb~Raha,~\IEEEmembership{Student Member,~IEEE,}
		Apurba~Adhikary,
		Walid Saad,~\IEEEmembership{Fellow,~IEEE,}
		Zhu~Han,~\IEEEmembership{Fellow,~IEEE,}
		and~Choong Seon~Hong,~\IEEEmembership{Fellow,~IEEE}
		\thanks{Mrityunjoy Gain is with the Department of Artificial Intelligence, School of Computing,
		Kyung Hee University, Yongin 17104, Republic of Korea. (e-mail: gain@khu.ac.kr).
		
		Kitae Kim, Avi Deb Raha, and Choong Seon Hong are with the Department of Computer Science and Engineering, School of Computing, Kyung Hee University, Yongin 17104, Republic of Korea. (e-mail: glideslope@khu.ac.kr; avi@khu.ac.kr; cshong@khu.ac.kr).
			
		Apurba Adhikary is with the Department of Information and Communication Engineering, Noakhali Science and Technology University, Noakhali-3814, Bangladesh (e-mail: apurba@khu.ac.kr)
		
		Walid Saad is with the Bradley Department of Electrical and Computer Engineering, Virginia Tech, Arlington, VA, 22203, USA, email(walids@vt.edu).
		
		Zhu Han is with the Electrical and Computer Engineering Department,
		University of Houston, Houston, TX 77004 (email: hanzhu22@gmail.com).
		
		Corresponding author: Choong Seon Hong (e-mail: cshong@khu.ac.kr).
	}
\vspace{-5mm}
}
	
	\markboth{Preprint submitted to IEEE Transactions for peer review}%
	{Shell \MakeLowercase{\textit{et al.}}: A Sample Article Using IEEEtran.cls for IEEE Journals}
	
	
	\maketitle
	
	\begin{abstract}
	The Artificial Intelligence Radio Access Network (AI-RAN) paradigm enhances network intelligence by integrating AI and ML into RAN operations and infrastructure, enabling dynamic optimization and real-time adaptability while supporting eMBB, uRLLC, mMTC, and emerging AI-native applications. While these services traditionally required dedicated devices, next-generation (NextG) networks aim to enable a single user equipment (UE) to support multiple services concurrently, critical for applications like the metaverse. In this paper, we propose a deep incremental framework for efficient RAN management, introducing the Multi-Service-Modal UE (MSMU) system, which enables a single UE to handle eMBB and uRLLC services simultaneously. We formulate an optimization problem integrating traffic demand prediction, route optimization, RAN slicing, service identification, and radio resource management under uncertainty. We decompose it into long-term (L-SP) and short-term (S-SP) subproblems then propose a Transformer model for L-SP optimization, predicting eMBB and uRLLC traffic demands and optimizing routes for RAN slicing. To address non-stationary network traffic with evolving trends and scale variations, we integrate reversible instance normalization (ReVIN) into the forecasting pipeline. For the S-SP, we propose an LSTM model enabling real-time service type identification and resource management, utilizing L-SP predictions. We incorporate continual learning into the S-SP framework to adapt to new service types while preserving prior knowledge. Experimental results demonstrate that our proposed framework achieves up to 46.86\% reduction in traffic demand prediction error, 26.70\% and 18.79\% improvement in PRBs and power estimation, 7.23\% higher route selection accuracy, and 7.29\% improvement in service identification over the baselines with 95\% average accuracy in continual service identification across seven sequential tasks. 
	\end{abstract}


	\section{Introduction}
	Next-generation (NextG) networks, including 5G, 6G and beyond, need to support diverse services like enhanced mobile broadband (eMBB), ultra reliable low latency communication (uRLLC), massive machine type communication (mMTC), holographic-type communications, integrated sensing and communication etc. each with unique requirements for throughput, reliability, and latency \cite{saad2, saad3, pacifista_TMC2, orchestration_TMC3}. The Open Radio Access Network (O-RAN) concept enhances network flexibility by disaggregating RAN components into the Radio Unit (RU), Distributed Unit (DU), and Centralized Unit (CU), while incorporating open, standardized interfaces. This enables vendor interoperability across hardware, software, and interfaces, fostering intelligent network adaptability and optimization \cite{Saad, Securing_TMC1, vr_TMC4, dt_TMC5, oranalliance}. However, O-RAN treats AI as a separate application, leading to inefficient resource use and limited real-time processing below 10ms, critical for uRLLC, and lacks native edge AI support \cite{aiml_oran2024}. The Artificial Intelligence Radio Access Network (AI-RAN) Alliance aim to address the limitations of O-RAN by integrating AI directly into the network core \cite{airanbriefing}. The AI-RAN focuses on three key goals: using AI for RAN to improve spectral efficiency, AI and RAN to converge workloads onto a shared platform, and AI on RAN to transform the network into a platform for new services \cite{airan2024whitepaper}.
	
	AI-RAN builds on O-RAN's three layer architecture management (non real time), control (near real time), and function (real time) by deeply integrating AI and machine learning. The management layer ($>$1s) handles AI model training, lifecycle management, and policy orchestration. The control layer (10ms–1s) enables AI driven real time radio resource management. Unlike O-RAN, AI-RAN embeds AI in the function layer ($<$10ms) for distributed, edge based tasks like scheduling and power control, enabling dynamic, autonomous, and context aware network optimization \cite{aiml_oran2024, airanbriefing, airan2024whitepaper}. Managing multiple different services like uRLLC, eMBB, and mMTC on a single RAN framework is a significant challenge, as constructing separate networks for each service type is impractical \cite{airan2024whitepaper}. Therefore, efficiently routing heterogeneous traffic to enhance user experience and network performance remains a challenge \cite{rost}.
	
	Numerous works have attempted to overcome the difficulties outlined above \cite{Wu, Anand, Zhang, Nguyen, Korrai, Swain, solmaz, Bonati, Kavehmadavani_ICCW,  pandora_TMC6, sla_TMC7, orchestran_TMC8}. In \cite{Wu}, the authors proposed a puncturing method to reduce uRLLC queuing delays. However, this approach lowers eMBB throughput when uRLLC traffic is high. The work in \cite{Anand} proposed a joint scheduling approach of uRLLC and eMBB traffic to minimize uRLLC resource usage while maximizing eMBB throughput to meet quality of service (QoS) requirements. Traffic steering (TS) and RAN slicing were further investigated in \cite{Zhang}, which evaluated a dynamic multi-connectivity (MC) approach. The authors in  \cite{Nguyen} analyzed mixed numerologies for scheduling heterogeneous services with varying QoS needs. In \cite{Korrai}, the authors examined combined optimization for radio resource slicing, resource block, and power assignment, while taking into account the presence of imperfect channel state information (CSI) in 5G networks. Finally, the work in \cite{Swain} proposed an AI-driven model for traffic management in 6G cloud RANs to improve resource allocation and network performance. However, the previously described studies have focused on TS with adaptive numerology within a uniform, ``one-size-fits-all" network architecture. This approach does not provide sufficient adaptability to accommodate the diverse requirements of heterogeneous services, such as eMBB, uRLLC, and mMTC. While O-RAN offers significant benefits, limited research has focused on applying it specifically to TS. For example, the work \cite{solmaz} introduced an intelligent framework for radio resource management and traffic prediction for congested cells in O-RAN, while the work in \cite{Bonati} analyzed the integration of intelligence at each O-RAN layer for data-driven NextG networks. In \cite{Kavehmadavani_ICCW}, the authors proposed a TS system that makes use of MC and RAN slicing with fixed numerology tailored to 5G NR. However, Many of these studies focus on fixed numerology within the inflexible 4G LTE architecture for resource allocation, which limits the ability to support heterogeneous traffic in beyond 5G networks. These works do not provide an integrated solution for traffic steering and RAN resource slicing in O-RAN architecture, leaving a gap in meeting the diverse traffic demands expected beyond 5G wireless networks.
	
	That said, a number of recent works \cite{Kavehmadavani_TWC1, VDN, Lacava_TMC, Kavehmadavani_Gcom, Kavehmadavani_TWC2, Sroka, Linsalata, Polese, Jingwen, Yeh, Marcin, Motalleb} developed intelligent TS and resource optimization approaches for O-RAN. In \cite{Kavehmadavani_TWC1}, the authors proposed an long short term memory (LSTM) based framework for RAN management and traffic prediction, with the goal of optimizing resource utilization based on unpredictable traffic demands. The work in \cite{VDN} introduced an adaptive TS framework using reinforcement learning and network utility maximization to enhance delay-utility tradeoffs in dynamic environments. The ns-O-RAN framework developed in \cite{Lacava_TMC}, combined a near real time RAN intelligent controller (near-RT RIC) with 3GPP-based simulations to support xApp development and user-specific TS policies. Additionally, the work in  \cite{Kavehmadavani_Gcom} introduced a deep reinforcement learning (DRL) based TS approach at the non real time RAN intelligent controller (non-RT RIC) for optimized multi-service downlink allocation. In \cite{Kavehmadavani_TWC2}, the authors integrated multi-layer optimization with LSTM and multi agent DRL for dynamic resource management across O-RAN’s timescales. In the context of vehicle-to-everything (V2X) communications, the work in \cite{Sroka} proposed the concept of O-RAN enrichment for adaptive policy control, while the work in \cite{Linsalata} developed stable multi-hop connectivity for mmWave-based connected and autonomous vehicles (CAVs), addressing vehicle-to-vehicle (V2V) communication and relay selection challenges. In \cite{Polese}, the authors discussed how O-RAN principles enhance 6G networks' flexibility and efficiency, promoting standardization efforts. In \cite{Jingwen}, the authors introduced an intent-driven framework for network slicing automation using deep reinforcement learning. \cite{Yeh} introduces an intelligent xApp for IoT services with service level agreement compliance, integrated with a near RT-RIC.  The work in \cite{Marcin} focused on xApp and rApp development for O-RAN, with application to beam management, while in the work in \cite{Motalleb} the authors focused on baseband resource allocation and virtualized network function (VNF) activation. Traditionally, RAN services like eMBB, uRLLC, and mMTC have been managed for separate user equipment. However, the emergence of applications such as Web 3.0 and the metaverse necessitates multi-service capabilities on a single device, which requires the simultaneous support of high data rates and low latency. While O-RAN and AI-RAN offer a foundation for such a capability, current RAN management and resource slicing schemes have not been optimized to efficiently support multiple services on a single UE. This underexplored area represents a critical challenge for future network architectures, thus necessitating focused research and development.
	
	In this paper, we propose a novel deep incremental AI-RAN management framework for optimizing resource slicing and allocation, including transmit power and time-frequency units. Conventional RAN management strategies often lead to inefficient resource utilization and statically assign dedicated UEs to a single service type. To overcome these limitations, we propose an adaptive mechanism, considering multi-service-modal UEs (MSMU) in the AI-RAN framework, that facilitates the dynamic coexistence of multiple traffic types, such as eMBB and URLLC, on a single UE. We then integrate MC with 3GPP Release 15 mixed numerologies to enable users to connect to multiple nodes using service-specific resource allocations, thereby improving throughput and reliability while reducing interference and latency for uRLLC and eMBB coexistence in bandwidth-limited networks. We take a proactive approach by first using a Transformer model to forecast traffic demand and select routes for both eMBB and uRLLC services. Based on these predictions, we apply a heuristic method for radio resource slicing, followed by an LSTM model to handle service identification, user association, power estimation, and final resource allocation. Given the non-stationary nature of network traffic, we integrate Reversible Instance Normalization (ReVIN) \cite{kim2022reversible} to stabilize input dynamics for focusing on meaningful temporal patterns, resulting in more accurate predictions. Aligned with AI-RAN standards, the framework optimizes eMBB throughput while ensuring uRLLC latency requirements are met. As new types of traffics may arise over time, we incorporate continual learning with the LSTM model to ensure seamless adaptation to new traffic patterns without forgetting previously learned data \cite{gain_noms, Chen, Benzaid}, enabling efficient resource management for emerging applications like Web 3.0. The main contributions of our proposed method are illustrated below.
	
	\begin{itemize}
		\item  We propose a novel deep incremental AI-RAN framework to support MSMU for seamless coexistence of eMBB and uRLLC services, introducing multi-service functionality on a single device. 
		
		\item  We develop an optimization framework for traffic demand forecasting, route prediction, RAN slicing, service identification, user association, and radio resource management while satisfying, power, and resource constraints.
		
		\item  To effectively address the formulated problem within the AI-RAN architecture, we decompose the problem into long-term (L-SP) and short-term (S-SP) subproblems, mapped to non-RT RIC rAPPs for traffic forecasting and RAN slicing, and near-RT RIC xApps for service type identification and resource management, with AI/ML-embedded CU for queue management.
		
		\item  For solving the L-SP, we propose a ReVIN empowered Transformer model to predict eMBB and uRLLC traffic demand and routing interfaces, and apply heuristic methods for RAN slicing optimization.
		
		\item  For S-SP, we propose an LSTM model to forecast service type, required physical resource blocks (PRBs), and power in real-time for each UE, integrating insights from non-RT RIC through the A1 interface. We incorporate continual learning in xApp1 (S-SP) to enable the model to adapt to new eMBB and uRLLC service types without forgetting previously learned types.
		
		
		\item Simulation results validate the proposed framework's effectiveness in managing MSMU, achieving up to a 46.86\% reduction in traffic demand prediction error, 26.70\% improvement in PRBs and power estimation, 7.23\% higher route selection accuracy, and 7.29\% improvement in service identification over baseline models. Furthermore, the framework demonstrates high predictive performance with over 99\% accuracy in service type and route prediction, alongside efficient continual learning performance exceeding 95\% accuracy across seven tasks.
	\end{itemize}
	
	This paper's remaining sections are arranged as follows. Section \ref{SM} provides an illustration of the system model. We outline the problem definition and the proposed solution approach in Sections \ref{P_F} and \ref{SA}, respectively. Section \ref{Ex} presents the simulation results and analysis, and finally Section \ref{Con} concludes the paper.

	\section{System Model} \label{SM}
	\subsection{AI-RAN Scenario}
\begin{figure}[t]
		\centerline{\includegraphics[width=1\linewidth, height = 9cm]{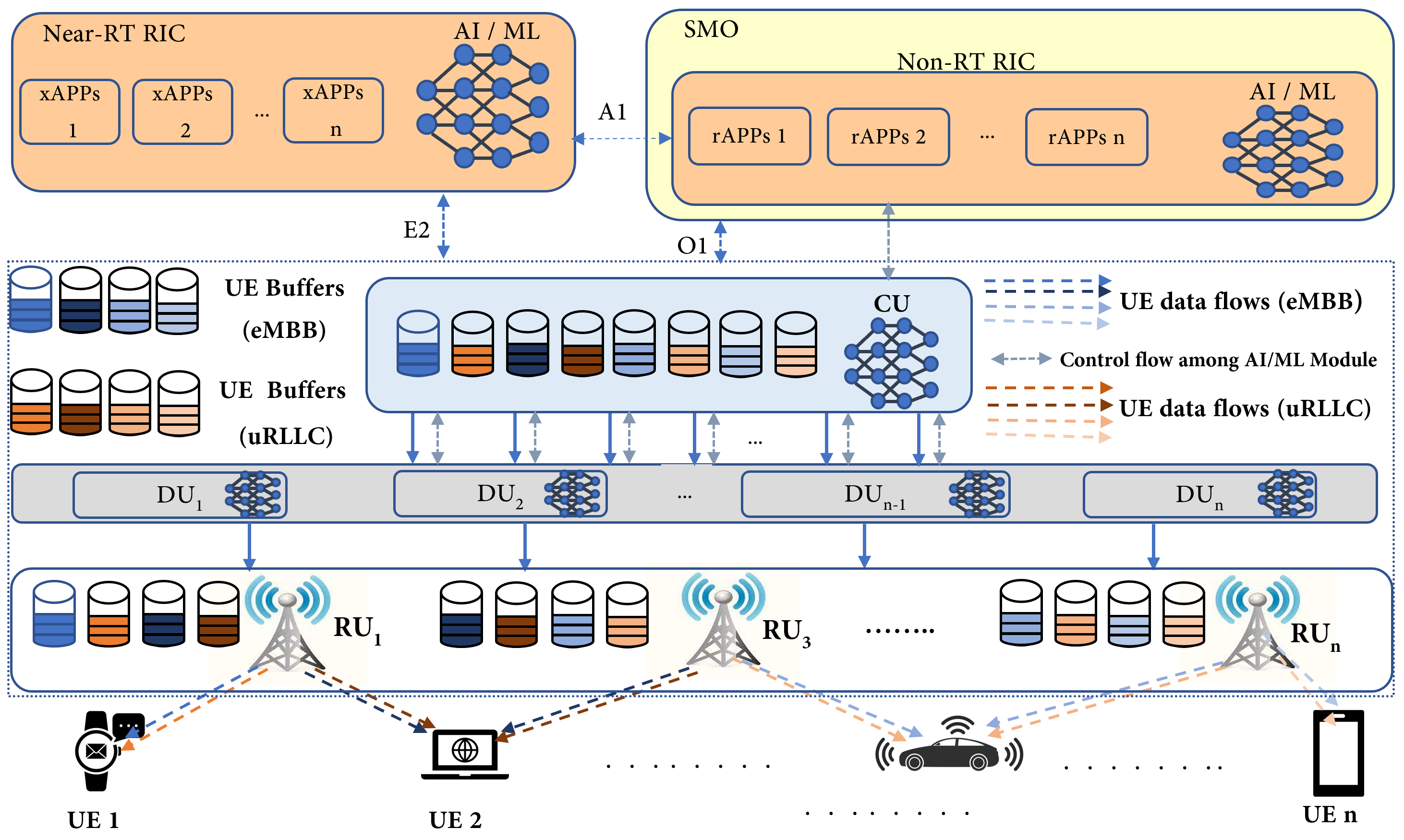}}
		\caption{System model for multi-service-modal UE (MSMU) management within the AI-RAN framework. Each UE supports both eMBB and uRLLC services, with data buffered at the RU, DU, and CU in device-service-specific buffers. Besides AI capabilities in the RIC (as in O-RAN), the system model embeds AI functionalities directly into network components such as the CU and DU.}
		\label{AI-RAN-system-model}
	\end{figure}

\begin{figure*}[t]
		\centerline{\includegraphics[width=.95\linewidth, height=8cm]{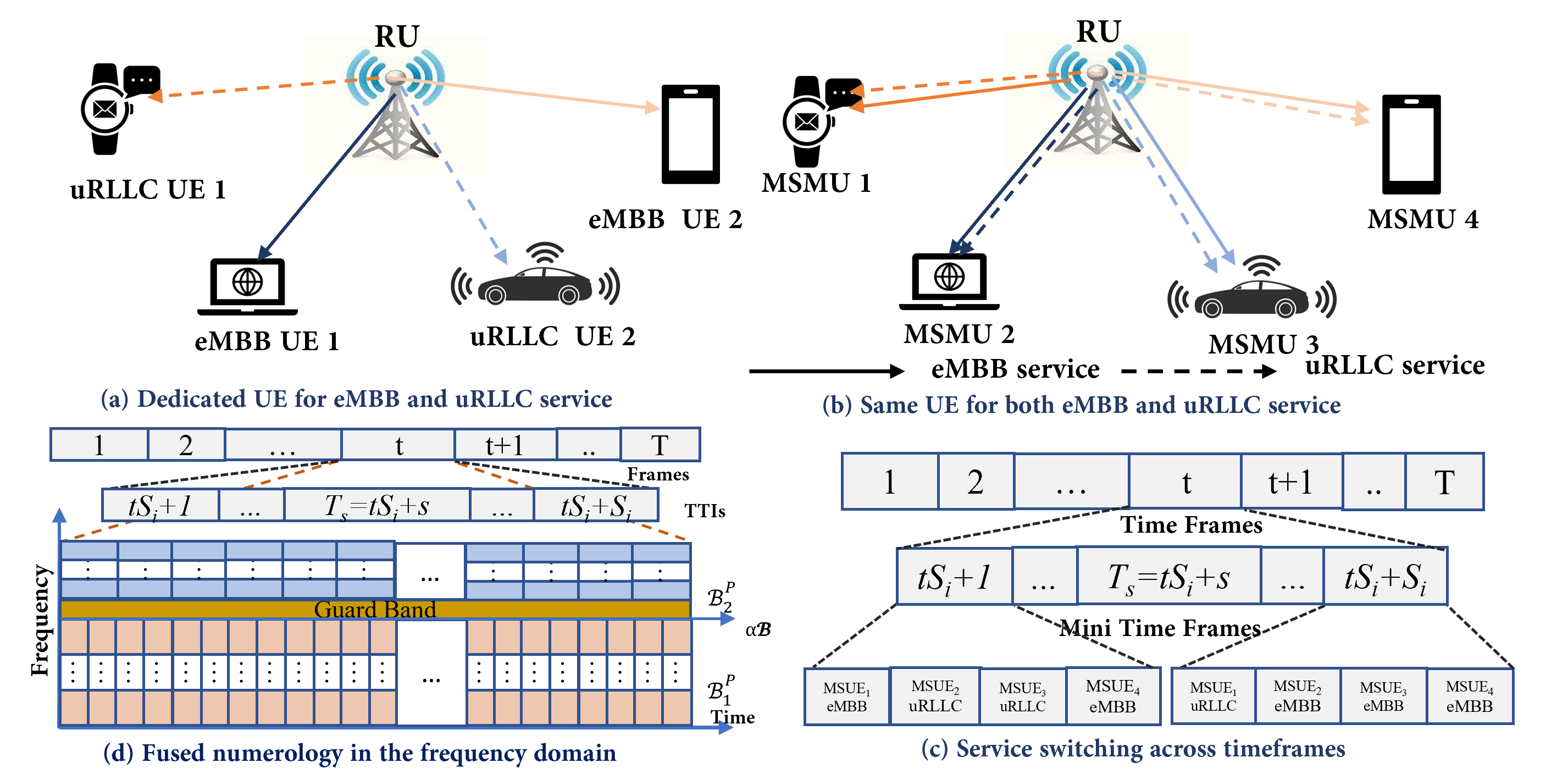}}
		\caption{proposed mixed numerology in frequency domain, multi-service-modal UE (MSMU) over service dedicated UE and proposed service switching mechanism. The proposed service-switching mechanism allows the same UE to dynamically switch between different services at each mini time frame, enabling support for multiple services in rapid succession.}
		\label{msmu_numer}
		\vspace{-3mm}
	\end{figure*}

	As illustrated in Fig. \ref{AI-RAN-system-model}, we consider a downlink orthogonal frequency-division multiple access (OFDMA) multi-user multiple-input single-output (MU-MISO) system within an AI RAN framework. The system comprises a single CU, a collection of DUs represented by the set $\mathcal{F} = \{1, 2, \ldots, F\}$, and a set of RUs denoted by $\mathcal{E} = \{1, 2, \ldots, E\}$. Each DU $f$ is responsible for serving a subset of RUs, denoted by $\mathcal{E}_f = \{(f, 1), \ldots, (f, E_f)\}$, where $|\mathcal{E}_f| = E_f$, and the total number of RUs in the network satisfies $\sum_{f \in \mathcal{F}} E_f = E$. Each RU $(f, e)$ is equipped with $K$ antennas, while each UE has a single antenna.
	The set of UEs is denoted by $\mathbb{U} = \{1, \ldots, u\}$, and each UE is served by the RUs. User services are classified into two types: eMBB and uRLLC, denoted by the subsets $\mathbb{S}_{\text{em}}$ and $\mathbb{S}_{\text{ur}}$ of the service set $\mathbb{S}$. A user may require either or both types of services. An AI-RAN system is designed to intelligently manage these heterogeneous services, which are defined as:
	\begin{align}
		\label{user_service_set}
		&\mathbb{S}_u = \mathbb{S}_{\text{em}} \cup \mathbb{S}_{\text{ur}}, \\
		\label{embb_set}
		&\mathbb{S}_{\text{em}} = \{\mathbb{S} \in \mathbb{S}_u \mid B(\mathbb{S}) > B_{\text{th}} \}, \\
		\label{urllc_set}
		&\mathbb{S}_{\text{ur}} = \{\mathbb{S} \in \mathbb{S}_u \mid L(\mathbb{S}) < L_{\text{th}} \},
	\end{align}
	where $\mathbb{S}_u$ is the set of all user services. $\mathbb{S}_{\text{em}}$ is the set of eMBB services. $\mathbb{S}_{\text{ur}}$ is the set of URLLC services. $B(\mathbb{S})$ and $L(\mathbb{S})$ are the bandwidth and latency of a service $\mathbb{S}$, respectively. $B_{\text{th}}$ and $L_{\text{th}}$ are the predefined thresholds for bandwidth and latency. The eMBB services transmit large packets of size $X^{\text{em}}$ bytes, while uRLLC services transmit smaller packets of size $X^{\text{ur}}$ bytes. Data is stored in per-user, per-service buffers until transmission. Frequency-time resource blocks (RBs) are assigned transmission power by the RUs, with AI-driven radio resource management and scheduling enabling dynamic, autonomous, and context-aware network optimization. The system operates over mini-slots, where each timeframe is divided into two mini-timeframes of duration $\delta_\gamma = \frac{1}{2^{\gamma+1}}$ ms. Each mini-timeframe consists of 7 OFDM symbols with a subcarrier spacing (SCS) determined by numerology index $\gamma$, where $\text{SCS} = 15 \times 2^{\gamma}$ kHz. We adopt a multi-numerology framework in which UEs can dynamically switch between eMBB and uRLLC services across consecutive mini-timeframes, as illustrated in Fig. \ref{msmu_numer}. Following \cite{inter_numerology}, we prioritize numerology $\gamma = 1$ for eMBB, which corresponds to an SCS of 30 kHz, RB bandwidth $\beta_1 = 360$ kHz, and transmission time interval (TTI) $\delta_1 = 0.25$ ms. For uRLLC, numerology $\gamma = 2$ is used, corresponding to an SCS of 60 kHz, RB bandwidth $\beta_2 = 720$ kHz, and TTI $\delta_2 = 0.125$ ms.

	In our model, we consider multiplexing mixed numerologies in the frequency domain, where the downlink carrier bandwidth is divided into multiple bandwidth parts (BWPs). Users adjust their RF bandwidth by alternating between BWPs according to the necessary data rates. As shown in Fig. \ref{msmu_numer}(c), BWPs are designed based on the queue length of each service, with the bandwidth-split variable $\Phi[t]$ in the range $[0, 1]$. To reduce inter-numerology interference (INI), a fixed guard band ($B^g = 180$ kHz) is placed between adjacent sub-bands. The uRLLC slice BWP with numerology index $\gamma = 2$ is defined as $B_{\gamma}[t]\big|_{\gamma=2} = \Phi[t] B$, while the eMBB slice BWP with numerology index $\gamma = 1$ is $B_{\gamma}[t]\big|_{\gamma=1} = (1 - \Phi[t]) B - B^g$.
	
	We assume the system operates within discrete time-frames indexed by $t$, ranging from 1 to $T$, corresponding to a large-scale coherence time $\Delta = 10$ ms per frame, as shown in Fig. \ref{msmu_numer}(c). Each frame is partitioned into $S_\gamma$ TTIs based on the selected numerology $\gamma$, with each TTI duration $\delta_\gamma$. Each BWP is divided into $O_\gamma$ sub-bands, where $O_\gamma[t] = \lfloor B_\gamma[t]/\beta_\gamma \rfloor$ and $S_\gamma = \Delta/\delta_\gamma$. Therefore, the total number of RBs available for numerology index $\gamma$ in each frame $t$ is $O_\gamma[t] \times S_\gamma$.
	
	As shown in Fig. \ref{AI-RAN-system-model}, $U$ independent user traffic flows are forwarded to VNFs in the DU layer for parallel processing in response to requests from the CU. To model the RU-side processing behavior, we adopt an $M/M/1$ queue, assuming Poisson arrivals and exponential service times, suitable for representing single-threaded computation where tasks are served on a first-come-first-served (FCFS) basis. This abstraction captures the stochastic nature of lightweight packet processing, particularly for uRLLC or real-time control tasks. Each user may utilize up to $E$ parallel paths between CU and RUs. Accordingly, the CU splits the user's data into sub-flows, which are transmitted over distinct fronthaul paths and later reassembled at the UE. Following the approach in \cite{Kavehmadavani_TWC1}, we define the flow-split selection vector $\mathbf{a}_u[t] \triangleq [a_{e,u}[t]]$, where $a_{e,u}[t] = 1$ indicates that RU $e$ is selected for data transmission, and $a_{e,u}[t] = 0$ otherwise. We define $\boldsymbol{R}[t] \triangleq \{\boldsymbol{R}_u[t], \forall u \mid \sum_{e} R_{e,u}[t] = 1, R_{e,u}[t] \in [0, 1]\}$ as the global routing decision, in which $\boldsymbol{R}_u[t] \triangleq \{R_{e,u}[t]\}_{T}$ represents the routing portion vector of user $u$, while $\sum_{e} R_{e,u}[t] = 1$, where $R_{e,u}[t] \in [0, 1]$ indicates a piece of the data flow is routed to user $u$ via RU $e$ by choosing act $a_{e,u}[t]$ in time $t$. Next, we present the wireless channel model along with the achievable throughput for eMBB services and the latency characteristics for uRLLC services for the proposed framework.
	
	\subsubsection{Wireless Channel Model} The channel vector between RU $e$ and user $u$ at the sub-band $o_i$ in TTI $t_s$ is given by $\boldsymbol{h}_{e,u}^{o_i}[t_s] \in \mathbb{C}^{K \times 1}$ in accordance with the Rician factor and the Rician fading model $\xi_{e,u}^{o_i}[t]$. While the channel may vary over each short-time scale TTI, we assume that it stays temporally invariant inside each frame. We model $\boldsymbol{h}_{e,u}^{o_i}[t_s]$ as follows:
	
	\begin{equation}
		\label{}
		\begin{aligned}
			\boldsymbol{h}_{e,u}^{o_i}[t_s] = & \sqrt{\chi_{e,u}^{o_i}[t]}(\sqrt{\xi_{e,u}^{o_i}[t]/({\xi_{e,u}^{o_i}[t] + 1}}) \boldsymbol{\bar{h}}_{e,u}^{o_i}[t] \\ & + \sqrt{1 / (\xi_{e,u}^{o_i}[t] + 1})\boldsymbol{\tilde{h}}_{e,u}^{o_i}[t_s]),
		\end{aligned}
		\label{tput}
	\end{equation}
	where $\chi_{e,u}^{o_i}[t]$ represents the large-scale fading, $\mathbf{\bar{h}}_{e,u}^{o_i}[t]$ and $\mathbf{\tilde{h}}_{e,u}^{o_i}[t_s]$ represent the non-LoS (NLoS) and line-of-sight (LoS) components, which, respectively, adhere to a Rayleigh fading model and a deterministic channel. The Rician fading model is chosen as it accurately captures the presence of a dominant LoS path in urban and suburban wireless environments while incorporating multipath scattering effects through the Rayleigh-distributed NLoS component.
	
	\subsubsection{Achievable Throughput}
	Given the orthogonality constraint, we assume that a single user is assigned to each RB of an RU during a single TTI. This assignment is represented by binary variables $\Psi^{\text{em},o_i}_{e,u}[t_s] \in \{0, 1\}$ and $\Psi^{\text{ur},o_i}_{e,u}[t_s] \in \{0, 1\}$ for eMBB and uRLLC traffic, respectively. In our system model, the service type of each UE is dynamic, with the possibility of requiring either eMBB ($\mathbb{S}_{\text{em}}$) or uRLLC ($\mathbb{S}_{\text{ur}}$) services. Each UE is allocated a single type of resource block per TTI, based on its service category. For each TTI $t_s$, sub-band $o_i$, and RU $m$, we define the RB assignment as follows:
	\[
	\Psi^{\text{em},o_i}_{e,u}[t_s] = 
	\begin{cases} 
		1 & \text{if } u \in \mathbb{S}_{\text{em}} \text{ and the RB } (t_s, o_i) \text{ is assigned} \\ 
		0 & \text{otherwise},
	\end{cases}
	\]
	\[
	\Psi^{\text{ur},o_i}_{e,u}[t_s] = 
	\begin{cases} 
		1 & \text{if } u \in \mathbb{S}_{\text{ur}} \text{ and the RB } (t_s, o_i) \text{ is assigned} \\ 
		0 & \text{otherwise}.
	\end{cases}
	\]
	
	Let $\Xi[t_s] \coloneqq \bigl\{ \left\{ \Psi^{\text{em},o_i}_{e,u}[t_s],\ \Psi^{\text{ur},o_i}_{e,u}[t_s] \right\} \allowbreak \mid \sum_{e,u} \Psi^{\text{em},o_i}_{e,u}[t_s] + \Psi^{\text{ur},o_i}_{e,u}[t_s] \leq 1,\ \allowbreak \Psi^{\text{em},o_i}_{e,u}[t_s],\ \Psi^{\text{ur},o_i}_{e,u}[t_s] \in \{0,1\}\ \allowbreak \bigr\}$ be the RB allocation constraint. For the uRLLC service, this requirement guarantees both orthogonality and QoS. Using the MC technique, interference on eMBB traffic is minimized, with residual interference treated as constant \cite{MC}. Thus, the achievable rate for the $u$-th eMBB user at TTI $t_s$ is:
	
	\begin{equation}
		\begin{aligned}
			\mathcal{R}^{\text{em}}_{e,u}(\boldsymbol{\mathcal{P}}^{\text{em}}[t_s])=\sum_{{o_i}=1}^{O_i} \beta_i \log_2 \left(1+ \frac{\mathcal{P}^{\text{em},o_i}_{e,u}[t_s] g_{e,u}^{o_i}[t_s]}{\sigma^2}\right),
		\end{aligned}
		\label{ar}
	\end{equation}
	where, $\beta_i$, $\sigma^2$, and $\mathcal{P}_{\text{em},o_i}^{e,u}[t_s]$ indicate the bandwidth of every RB in numerology $i$, the additive white Gaussian noise power, and transmission power from RU $e$ to user $u$ for TTI $t_s$ eMBB traffic at sub-band $o_i$, respectively. $g_{e,u}^{o_i}[t_s]$ represents the effective channel gain, given by $g_{e,u}^{o_i}[t_s] \triangleq \left\| \boldsymbol{h}_{e,u}^{o_i}[t_s] \right\|^2_2$. We define $\boldsymbol{\mathcal{P}}^{\text{em}}[t_s] \triangleq \left[ \mathcal{P}^{\text{em},o_i}_{e,u}[t_s] \right]$ for all $o_i$, $u$, and $e$. 		
	The transmit power must satisfy the constraint $\sum_{u \in \mathbb{S}_{\text{em}}} \sum_{o_i} \mathcal{P}^{\text{em},o_i}_{e,u}[t_s] \leq \mathcal{P}^{\text{max}}_e$ to ensure that the total transmit power of RU $e$ across all users and RBs in TTI $t_s$ does not exceed its power budget. Additionally, to enforce valid RB assignments, each individual allocation must satisfy $\mathcal{P}^{\text{em},o_i}_{e,u}[t_s] \leq \Psi_{e,u}^{\text{em},o_i}[t_s] \mathcal{P}^{\text{max}}_e$, where $\Psi_{e,u}^{\text{em},o_i}[t_s]$ is a binary variable indicating whether user $u$ is assigned to RB $o_i$ at RU $e$ in TTI $t_s$. Consequently, the throughput of eMBB user $u \in \mathbb{S}_{\text{em}}$ in TTI $t_s$ is given as $\mathcal{R}^{\text{em}}_{u}(\boldsymbol{\mathcal{P}}^{\text{em}}[t_s]) = \sum_e \mathcal{R}_{e,u}^{\text{em}}(\boldsymbol{\mathcal{P}}^{\text{em}}[t_s])$.
	
	The constraint $\sum_{t_s} \mathcal{R}^{\text{em}}_{u}(\boldsymbol{\mathcal{P}}^{\text{em}}[t_s]) \geq \mathcal{R}_{\text{th}}$ ensures that eMBB users meet the minimum QoS requirement, where $\mathcal{R}_{\text{th}}$ is a predefined QoS threshold. In contrast, the instantaneous achievable rate of uRLLC user $u$ from RU $e$ in TTI $t_s$ utilizing the short block-length will be determined by:
	
	\begin{equation}
		\begin{aligned}
			\mathcal{R}^{\text{ur}}_{e,u}(\boldsymbol{\mathcal{P}}^{\text{ur}}[t_s], \boldsymbol{\Psi}^{\text{ur}}[t_s]) =  \sum_{o_i=1}^{O_i} \beta_i h \log_2 \bigg(1 + \mathcal{P}^{\text{ur},o_i}_{e,u}[t_s]\\ \frac{g_{e,u}^{o_i}[t_s]}{\sigma^2} \bigg) - \Psi_{e,u}^{\text{ur},o_i}[t_s] \sqrt{V} Q^{-1}(P_e) \sqrt{\frac{1}{\delta_i \beta_i}},
		\end{aligned}
	\end{equation}
	where $P_e$, $V$, and $Q^{-1}$ represent the error probability, channel dispersion, and inverse of the Gaussian Q-function, respectively. We define $\boldsymbol{\mathcal{P}}^{\text{ur}}[t_s] \triangleq \left[ \mathcal{P}^{\text{ur},o_i}_{e,u}[t_s] \right]$ and $\boldsymbol{\Psi}^{\text{ur}}[t_s] \triangleq \left[ \Psi_{e,u}^{\text{ur},o_i}[t_s] \right]$ for all $o_i$, $u$, and $e$. The power constraint is:
	
	\begin{equation}
		\label{eq:power_constraints}
		\begin{aligned}
			\mathscr{P}[t_s] \hspace{4mm}=\hspace{4mm} \{ 0 \leq \mathcal{P}^{\text{em},o_i}_{e,u}[t_s] \leq \Psi_{e,u}^{\text{em},o_i}[t_s] \mathcal{P}^{\text{max}}_e, \\
			\frac{\sigma^2 \Gamma_0 \Psi_{e,u}^{\text{ur},o_i}[t_s]}{g_{e,u}^{o_i}[t_s]} \leq \mathcal{P}^{\text{ur},o_i}_{e,u}[t_s] \leq \Psi_{e,u}^{\text{ur},o_i}[t_s] \mathcal{P}^{\text{max}}_e, \\
			\sum_i \sum_{o_i,u} (\mathcal{P}^{\text{em},o_i}_{e,u}[t_s] + \mathcal{P}_{\text{ur},o_i}^{e,u}[t_s]) \leq \mathcal{P}^{\text{max}}_e \}.
		\end{aligned}
	\end{equation}
	
	Lastly, we define $\Omega^{\text{em}}_u[t]$, and $\Omega^{\text{ur}}_u[t]$ (in packets per second) representing the user $u$'s unknown eMBB and uRLLC traffic demand, respectively, in time-frame $t$ (length $Z^\mathbb{S}$ bytes)., which is identical and independently distributed over time and with a finite constant $\Omega^{\text{max}}$ as the upper bound. The distinct queue that is kept for the $u$-th user for every service at each RU, given by $\{ R_{e,u}[t] \Omega_u[t] Z^\mathbb{S} \}$, symbolizes the sub-flow arriving procedures that are managed by a congestion scheduler. Consequently, the data flow's queue-length of $u$ at RU $e$ in TTI $t_{s+1}$ is given by $q^\mathbb{S}_{e,u}[t_{s+1}] = \max \{ q^\mathbb{S}_{e,u}[t_s] + R_{e,u}[t] \Omega_u[t] Z^\mathbb{S} \Delta - r_{e,u}^\mathbb{S}[t_{s+1}] \delta_i, 0 \}$. The constraint $\sum_u q^\mathbb{S}_{e,u}[t_s] \leq \mathcal{Q}^{\text{max}}$ is imposed to ensure queue stability, which means that the queue length stays bounded over time, and to prevent packet loss due to buffer overflow. This restricts the number of available packets in the RU's buffer from exceeding the maximum queue-length of $\mathcal{Q}^{\text{max}}$ for each RU. We define $\boldsymbol{q}^\mathbb{S}[t_s] \triangleq [q^\mathbb{S}_{e,u}[t_s]]^T, \forall e, u$.
	
	\subsubsection{Latency for uRLLC}
	In uRLLC services, meeting stringent delay and reliability constraints is critical. The total end-to-end (E2E) latency experienced by a uRLLC user includes processing delays at various network layers and transmission delays across transport and access links. We model this latency over discrete time-frames indexed by $t$, assuming a centralized RAN architecture comprising a CU, multiple DUs, and RUs.
	
	We denote $\eta_{\text{cu}}$ and $\eta_f$ the processing capacities (in CPU cycles per second) of the CU and DU, respectively, and $\mathcal{C}$ represent the number of cycles required to process one packet of size $\mathcal{Z}$ bits. The packet arrival rate for user $u$ at time $t$ is denoted by $\Omega_u[t]$, and the aggregated system arrival rate is $\Omega[t] = \sum_u \Omega_u[t]$. The midhaul and fronthaul links connecting the CU to DU and DU to RUs have respective capacities $\mathcal{C}_{\text{MH}}$ and $\mathcal{C}^{e}_{\text{FH}}$. The RU-to-UE (radio) interface allocates transmission resources dynamically based on instantaneous scheduling. The total E2E latency for uRLLC user $u$ is expressed compactly as:
	
	\begin{equation}
		\begin{aligned}
			\Upsilon_u^{\text{ur}}[t] &= 
			\frac{\sum_u \Omega_u[t] \mathcal{C}}{\eta_{\text{cu}}} +
			\frac{\sum_u \Omega_u[t] \mathcal{Z}}{\mathcal{C}_{\text{MH}}} +
			\frac{\sum_u \Omega_u[t] \mathcal{C}}{\eta_f} \\
			&\quad +
			\max_{e \in \mathcal{E}_f} \left( \frac{\sum_u R_{e,u}[t] \Omega_u[t] Z^{\text{ur}}}{\mathcal{C}^{e}_\text{FH}} \right) \\
			&\quad +
			\sum_{t_s} \left( 
			\max_{e \in \mathcal{E}_f} \left( \frac{R_{e,u}[t] \Omega_u[t] Z^{\text{ur}}}{r_{e,u}^{\text{ur}}[t_s]} \right) 
			+ \Upsilon^{\text{pro}}_e[t_s] 
			\right).
		\end{aligned}
	\end{equation}
		
	The first three terms capture delays introduced by centralized and distributed processing at the CU and DU, and by packet forwarding over the midhaul link. The fourth term models the fronthaul transmission delay, which may vary across RUs depending on their utilization and link capacity. The final term represents the combined delay from RU processing and radio access transmission over subframes $t_s$, where $r_{e,u}^{\text{ur}}[t_s]$ is the allocated air-interface rate, and $\Upsilon^{\text{pro}}_e[t_s]$ is the RU-side processing latency, typically spanning a small number of OFDM symbols. To ensure uRLLC reliability, the system imposes a probabilistic latency constraint of the form:
	
	\begin{equation}
		\Pr\left( \Upsilon_u^{\text{ur}}[t] \leq D_{\text{ur}} \right) \geq \epsilon_2,
	\end{equation}
	where $D_{\text{ur}}$ is the maximum allowable latency (e.g., 1 ms), and $\epsilon_2$ represents the target reliability, typically exceeding $99.999\%$. This constraint ensures that uRLLC services can meet their strict QoS guarantees even under dynamic traffic and scheduling conditions.

	\section{Problem Formulation} \label{P_F}
	Our objective is to optimize intelligent traffic prediction, route selection, RAN slicing, continual service identification, and radio resource management to efficiently serve the MSMU under diverse, conflicting requirements of eMBB and uRLLC services with dynamic traffic demands. The utility function must therefore capture the critical trade-off between the worst-case E2E uRLLC latency and the eMBB throughput. Specifically, this involves maximizing the eMBB throughput $ \mathscr{R}^{\text{em}} = \sum_{u \in \mathbb{S}_{\text{em}}} \mathcal{R}^\text{em}_u(\boldsymbol{\mathcal{P}}^{\text{em}}[t_s]) $ and minimizing the worst-user uRLLC latency $ \max_{u \in \mathbb{S}_{\text{ur}}} \{ \Upsilon^\text{ur}_u \} $ for MSMU. Consequently, the MSMU management problem is formulated as a multi-objective optimization task. To find the pareto optimal boundary that balances these conflicting goals, we apply the weighted sum method (WSM) to scalarize the problem, resulting in the following single-objective maximization problem:
	\begin{subequations}\label{p1_ran}
		\renewcommand{\theequation}{\theparentequation\alph{equation}} 
		\begin{align}
			\label{p1_ran:objective}
			& \hspace{-8mm}  \underset{\mathbb{S}, \boldsymbol{\Omega}^\textbf{em}, \boldsymbol{\Omega}^\textbf{ur}, \boldsymbol{\Psi}, \boldsymbol{R}, \boldsymbol{\mathcal{P}}, \Phi}{\text{max}}
			\quad \Big( \frac{\mathscr{R}^{\text{em}}(\boldsymbol{\mathcal{P}}^{\text{em}}[t_s])}{\mathcal{R}^{\text{max}}} - \lambda \cdot \frac{\max\{\Upsilon^{\text{ur}}_u\}}{D_{\text{ur}}} \Big), \tag{10}  \\ 
			\textbf{s. t.} \vspace{-2mm}
			&  \quad \mathbb{S}[t_s] \in \mathbb{S}_u, \label{p1_ran:const1} \\
			&  \quad \boldsymbol{\mathcal{P}}[t_s] \in \mathscr{P}[t_s], \; \forall t_s, \label{p1_ran:const2} \\
			&  \quad \boldsymbol{\Psi}[t_s] \in \Xi[t_s], \; \forall t_s, \label{p1_ran:const3} \\
			&  \quad \boldsymbol{R}_u[t] \in R[t], \; \forall u \in \mathbb{U}, \label{p1_ran:const4} \\
			&  \quad \hspace{-2mm} \text{Pr} \Bigl(\sum_{t_s} \mathcal{R}^{\text{em}}_u(\boldsymbol{\mathcal{P}}^{\text{em}}[t_s]) \geq \mathcal{R}_{\text{th}} \Bigr) \geq \epsilon_1, \; \forall u \in \mathbb{S}_{\text{em}}, \label{p1_ran:const5} \\
			&  \quad \sum_u \mathcal{R}_{e,u}^{\text{em}}(\boldsymbol{\mathcal{P}}^{\text{em}}[t_s]) 
			+ \mathcal{R}_{e,u}^{\text{ur}}(\boldsymbol{\mathcal{P}}^{\text{ur}}[t_s], \boldsymbol{\Psi}^{\text{ur}}[t_s]) \leq \mathcal{C}_e^{\text{FH}}, \notag \\  
			&  \hspace{55mm} \forall e \in E_f, \label{p1_ran:const6} \\
			&  \quad \sum_{t_s} \mathcal{R}^{\text{ur}}_{e,u} (\boldsymbol{\mathcal{P}}^{\text{ur}}[t_s], \boldsymbol{\Psi}^{\text{ur}}[t_s])
			\geq \frac{R_{e,u}[t] \Omega^\text{ur}_u[t] Z^{\text{ur}}}{\Delta}, \notag \\
			& \hspace{42mm} \forall e \in E_f, \; u \in \mathbb{S}_{\text{ur}}, \label{p1_ran:const7} \\
			&  \quad \text{Pr}(\Upsilon^{\text{ur}}_u(\boldsymbol{\Omega}^\text{ur}[t], \boldsymbol{R}[t], \boldsymbol{\Psi}[t_s], \boldsymbol{\mathcal{P}}[t_s]) \leq D_{\text{ur}}) \geq \epsilon_2, \; \notag \\ 
			&  \hspace{55mm} \forall u \in \mathbb{S}_{\text{ur}}, \label{p1_ran:const8} \\
			&  \quad \sum_u q^\mathbb{S}_{e,u}[t_s] \leq Q^\mathbb{S}_{\text{max}}, \; \forall t_s, \; e \in E_f, \label{p1_ran:const9} \\
			&  \quad \sum_{{o_i} = 1}^{O_i}  \beta_i \leq B_i[t], \; i \in \{1, 2\}, \label{p1_ran:const10} \\
			&  \quad 0 \leq \Phi[t] \leq 1, \label{p1_ran:const11}\\
			&  \quad \mathbb{S}_\text{new} \in \mathbb{S}_\text{em} \cup \mathbb{S}_\text{ur}, \forall \mathbb{S}_\text{new}, \label{p1_ran:const12}
		\end{align}
	\end{subequations}
	where $\lambda$ is a non-negative weighting factor ($\lambda \in [0, \infty)$), used to scalarize the two objectives. By sweeping $\lambda$, we obtain the entire pareto optimal boundary that captures the trade-off between the normalized eMBB throughput $\frac{\mathscr{R}^{\text{em}}}{\mathcal{R}^{\max}}$ and the normalized worst-user uRLLC latency $\frac{\max\{\Upsilon^{\text{ur}}_{u}\}}{D_{\text{ur}}}$. $ \mathcal{R}^{\max}$ is the maximum achievable eMBB throughput. The terms $\mathbb{S}[t]$, $\boldsymbol{\mathcal{P}}[t_s]$, $\boldsymbol{\Psi}[t_s]$, and $\boldsymbol{R}[t]$ from constraints (\ref{p1_ran:const1}), (\ref{p1_ran:const2}), (\ref{p1_ran:const3}), and (\ref{p1_ran:const4}) are the vectors encompassing the service type categorization, power allocation, and sub-band assignments and routes selection variables respectively. Constraint (\ref{p1_ran:const5}) indicates that the probability of the throughput for the eMBB services to touch a minimum threshold should be greater or equal to a certain positive threshold. The FH link between DU $f$ and RU $e$ has a restricted capacity, which is captured by constraint (\ref{p1_ran:const6}). Every RB issued to the uRLLC user $u$ is guaranteed to send a complete data packet of size $Z^{ur}$ by constraint (\ref{p1_ran:const7}). Constraints (\ref{p1_ran:const8}), and (\ref{p1_ran:const9}) capture the latency requirement for uRLLC services and, queue capacity for every service for every UE respectively. The probability that the delay $\Upsilon^{\text{ur}}_u$ does not exceed the maximum allowable delay $D_{\text{ur}}$ should be no less than a given positive threshold $\epsilon$. This probabilistic constraint accounts for the randomness and fluctuations in the arrival rate while ensuring a predefined level of performance. Constraints (\ref{p1_ran:const10}), and (\ref{p1_ran:const11}) capture the resource block bandwidth should be limited to the time frame bandwidth, and the slicing variable for each timeframe should lie within 0-1 respectively. Constraint (\ref{p1_ran:const12}) relates to the continual adaption of new services without forgetting the previous service knowledge.
	\begin{figure}[!t]
		\centerline{\includegraphics[width=\linewidth, height=3.5cm]{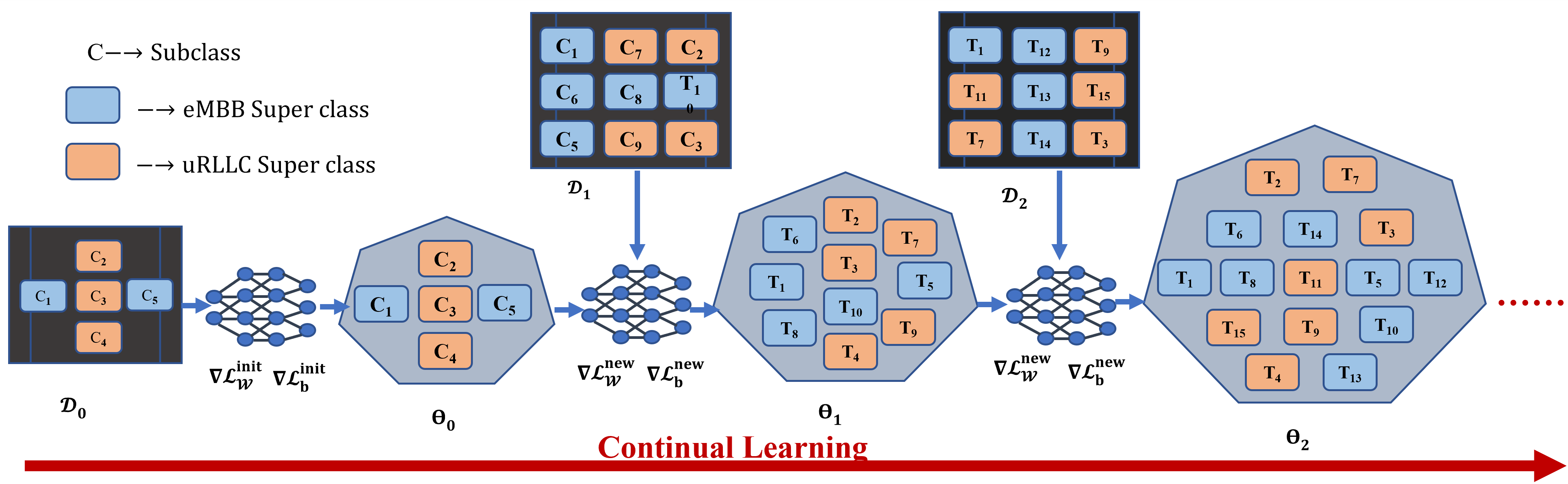}}
		\caption{System model illustrating the continual arrival and adaptation of sub-services under eMBB and uRLLC superclasses. As new sub-services such as video streaming, autonomous driving, cloud gaming, remote surgery etc. emerge over time, they must be accurately classified under the appropriate superclass to enable proper network policy enforcement. Continual learning supports this process by adapting to new service patterns without forgetting previously learned classes.}
		\label{cl}
	\end{figure}

	\subsection{Proposed Decomposition of Problem (\ref{p1_ran})}
	The main challenges in solving the formulated problems in (\ref{p1_ran}) lie in the nonconvexity of $\Upsilon_{u}^\text{ur}$ and constraints (\ref{p1_ran:const6}), (\ref{p1_ran:const7}), and (\ref{p1_ran:const9}) with respect to route variables and transmission power variables. Furthermore, these issues are typically mixed-integer nonlinear convex programming (MINCP) problems, which are more challenging to solve straight due to the binary structure of the sub-band allotment variables in constraint (\ref{p1_ran:const3}). Therefore the problem is NP-hard and no polynomial-time algorithm is known to guarantee finding the global optimum. Moreover, the traffic demand $\boldsymbol{\Omega}^\text{em}[t]$ and $\boldsymbol{\Omega}^\text{ur}[t]$ for the upcoming time frame are uncertain in real time. Thus, the bandwidth split $\Phi[t]$ for timeframe $t$ will be determined by the RAN layer's updated state and the traffic demands and routes $[\boldsymbol{\Omega}^\text{em}[t], \boldsymbol{\Omega}^\text{ur}[t], \boldsymbol{R}[t]]$. As a result, we suggest approaching the problem from two different time scales: the full time frame $t$ and the mini time frame $t_s$. The traffic demand variables $\boldsymbol{\Omega}^\text{em}[t]$ and $\boldsymbol{\Omega}^\text{ur}[t]$, the routing decision vector $\boldsymbol{R}[t]$, and the bandwidth-splitting variable $\Phi[t]$ are only resolved and upgraded once every time-frame $t$ to offer a robust queue structure and reduce computing burden and information exchange. In contrast, the service identification variable $\mathbb{S}[t_s]$ and corresponding RB allocation vector $\boldsymbol{\Psi}[t_s]$, and power assignment vector $\boldsymbol{\mathcal{P}}[t_s]$ are tailored to dynamic environments and adjusted in each mini time frame $t_s$. For appropriate SNR ranges, the uRLLC rate has the same concavity as the eMBB rate in (\ref{tput}) when the channel dispersion $V$ in (\ref{ar}) is approximated as 1. Next, we propose solution development for the problem in (\ref{p1_ran}).
	\subsubsection{Long-term Subproblem (L-SP)}
	At time-scale $t$, the traffic demand, routes, and dynamic RAN slicing joint optimization subproblem is reformulated as follows:
	\begin{subequations}\label{l-sp}
		\renewcommand{\theequation}{\theparentequation\alph{equation}} 
		\begin{align}
			\text{L-SP} :
			& \hspace{-1mm} \underset{\boldsymbol{\Omega}^\textbf{em}, \boldsymbol{\Omega}^\textbf{ur}, \boldsymbol{R}, \Phi}{\text{max}}
			\quad \Big( \frac{\mathscr{R}^{\text{em}}(\boldsymbol{\mathcal{P}}^{\text{em}}[t_s])}{\mathcal{R}^{\text{max}}} - \lambda \cdot \frac{\max\{\Upsilon^{\text{ur}}_u\}}{D_{\text{ur}}} \Big), \tag{11} \label{l-sp:objective} \\ 
			\hspace{-5mm} \textbf{s. t.} 
			&  \quad \boldsymbol{R}_u[t] \in R[t], \; \forall t,u, \label{l-sp:const1} \\
			&  \quad \sum_{t_s} \mathcal{R}^{\text{ur}}_{e,u}(\boldsymbol{\mathcal{P}}^{\text{ur}}[t_s], \boldsymbol{\Psi}^{\text{ur}}[t_s])
			\geq \frac{R_{e,u}\Omega^\text{ur}[t] Z^{\text{ur}}}{\Delta}, \notag \\
			& \hspace{40mm} \forall e \in E_f, \; u \in \mathbb{S}_{\text{ur}}, \label{l-sp:const2} \\
			&  \quad \text{Pr}(\Upsilon^{\text{ur}}_u(\boldsymbol{\Omega}^\text{ur}[t], \boldsymbol{R}[t], \boldsymbol{\Psi}[t_s], \boldsymbol{\mathcal{P}}[t_s]) \leq D_{\text{ur}}) \geq \epsilon_2, \; \notag \\
			& \hspace{50mm} \forall u \in \mathbb{S}_{\text{ur}}, \label{l-sp:const3} \\
			&  \quad \sum_{{o_i} = 1}^{O_i}  \beta_i \leq B_i[t], \; i \in \{1, 2\}, \label{l-sp:const4} \\
			&  \quad 0 \leq \Phi[t] \leq 1. \label{l-sp:const5}
		\end{align}
	\end{subequations}
	
	Since $\boldsymbol{\Omega}^\text{em}[t]$, $\boldsymbol{\Omega}^\text{ur}[t]$, and $\boldsymbol{R}[t]$ are all fully unknown at the start of each frame, problem in (\ref{l-sp:objective}) cannot be solved directly using standard optimization techniques, even though the objective function in (\ref{l-sp:objective}) is non-convex due to the nonconvexity of constraints (\ref{l-sp:const2}) and (\ref{l-sp:const3}). Therefore, in the future section, we propose AI-based methods to address the problem. Specifically, at the beginning of each frame $t$, we determine the traffic demand for both eMBB and uRLLC, dynamic bandwidth-split distribution, and dynamic routes variable as $\boldsymbol{\Omega}^\text{em}_*[t]$, $\boldsymbol{\Omega}^\text{ur}_*[t]$, $\boldsymbol{\Phi}^*[t]$, $\boldsymbol{R}^*[t]$, respectively. 
	
	\subsubsection{Short-term Subproblem (S-SP)}
	Given the parameters $\boldsymbol{\Omega}^\text{em}_*[t]$, $\boldsymbol{\Omega}^\text{ur}_*[t]$, $\boldsymbol{\Phi}^*[t]$, and $\boldsymbol{R}^*[t]$, received from the non-RT RIC via the A1 interface, the resource allocation problem at time slot $t_s$ within the near-RT RIC is formulated as:

	\begin{subequations}\label{s-sp}
		\renewcommand{\theequation}{\theparentequation\alph{equation}} 
		\begin{align}
			\text{S-SP} : 
			& \hspace{2mm}  \underset{\mathbb{S}, \boldsymbol{\Psi}, \boldsymbol{\mathcal{P}}}{\text{max}}
			\quad \Big( \frac{\mathscr{R}^{\text{em}}(\boldsymbol{\mathcal{P}}^{\text{em}}[t_s])}{\mathcal{R}^{\text{max}}} - \lambda \cdot \frac{\max\{\Upsilon^{\text{ur}}_u\}}{D_{\text{ur}}} \Big), \tag{12} \label{s-sp:objective} \\ 
			\textbf{s. t.} 
			&  \quad \mathbb{S}[t_s] \in \mathbb{S}_u, \label{s-sp:const1} \\
			&  \quad \boldsymbol{\Psi}[t_s] \in \Xi[t_s], \; \forall t_s, \label{s-sp:const2} \\
			&  \quad \boldsymbol{\mathcal{P}}[t_s] \in \mathscr{P}[t_s], \forall t, \label{s-sp:const3} \\
			& \hspace{-1mm} \quad \text{Pr} \Bigl(\sum_{t_s} \mathcal{R}^{\text{em}}_u(\boldsymbol{\mathcal{P}}^{\text{em}}[t_s]) \geq \mathcal{R}_{\text{th}} \Bigr) \geq \epsilon_1, \; \forall u \in \mathbb{S}_{\text{em}}, \hspace{-2mm} \label{s-sp:const4} \\
			&  \quad \sum_u \mathcal{R}_{e,u}^{\text{em}}(\boldsymbol{\mathcal{P}}^{\text{em}}[t_s]) 
			+ \mathcal{R}_{e,u}^{\text{ur}}(\boldsymbol{\mathcal{P}}^{\text{ur}}[t_s], \boldsymbol{\Psi}^{\text{ur}}[t_s]) \notag \\  
			&  \hspace{40mm} \leq \mathcal{C}_e^{\text{FH}}, \hspace{2mm} \forall e \in E_f, \label{s-sp:const5} \\
			&  \quad \sum_u q^\mathbb{S}_{e,u}[t_s] \leq Q^\mathbb{S}_{\text{max}}, \; \forall t_s, \; e \in E_f, \label{s-sp:const6} \\
			&  \quad \text{Pr}(\Upsilon^\text{ur}_u(\boldsymbol{\Psi}[t_s], \boldsymbol{\mathcal{P}}[t_s]) \leq D_{\text{ur}}) \geq \epsilon_2, \;  \forall u \in \mathbb{S}_{\text{ur}}, \label{s-sp:const7} \\
			&  \quad \sum_{t_s} \mathcal{R}^{\text{ur}}_{e,u}(\boldsymbol{\mathcal{P}}^{\text{ur}}[t_s], \boldsymbol{\Psi}^{\text{ur}}[t_s])
			\geq \frac{R_{e,u}^* \Omega^\text{ur}_{*u}[t] Z^{\text{ur}}}{\Delta}, \notag \\
			& \hspace{40mm} \forall e \in E_f, \; u \in \mathbb{S}_{\text{ur}}, \label{s-sp:const8} \\
			&  \quad \mathbb{S}_\text{new} \in \mathbb{S}_\text{em} \cup \mathbb{S}_\text{ur}, \forall \mathbb{S}_\text{new}. \label{s-sp:const9}
		\end{align}
	\end{subequations}
	
	The objective function in (\ref{s-sp:objective}) entails binary ($\Psi$) optimization variables at time slot $t_s$, which is still a MINCP problem, with a nonlinear objective function and non-convex constraint in (\ref{s-sp:const6}). MINCP problems constitute a vast class of challenging optimization problems because they include the difficulties of controlling nonlinear functions while optimizing under integer variables. We go through AI-based solutions to this issue in the following section. Constraint (\ref{s-sp:const9}) for the AI model is discussed in the context of continual learning and is defined as continual learning for the short-term subproblem (CL-S-SP) in the following subsection. This approach helps avoid unnecessary complexity or ambiguity, making both the S-SP and the mechanism behind the constraint easier to understand.
	\subsubsection{Continual Learning S-SP (CL-S-SP)}
	We consider dual-mode UEs supporting both eMBB and uRLLC services, with a focus on optimizing resource allocation based on application-specific service requirements. We consider two superclasses: eMBB for services like video streaming and file sharing, and uRLLC for services like autonomous vehicles and remote surgery etc. With the emergence of advanced technologies such as 6G and the metaverse, new categories of network services and application requirements are expected to arise. To address this, AI models must not only accurately classify services under the correct superclasses but also adapt to newly introduced subclasses over time. However, conventional deep learning models typically suffer from catastrophic forgetting, where learning new information leads to the loss of previously acquired knowledge. To overcome this limitation, we adopt continual learning strategy, which enables the model to incorporate new knowledge incrementally while retaining prior learning.
	
	We consider a sequence of tasks $\tau = (0, 1, 2, ..., \mathcal{T})$, each with data points that may contain new or old subclasses. Initially, dataset $\mathcal{D}_0$ consists of $\mathcal{N}_0$ samples with corresponding class labels, split into two superclasses ($\mathbb{S}_{\text{em}}$ and $\mathbb{S}_{\text{ur}}$). The model is trained on $\mathcal{D}_0$ to achieve good performance. When new data arrives at task $\tau$, we denote it as $\mathcal{D}_\tau$, which contains $\mathcal{N}_\tau$ samples of multiple subclasses belongs to eMBB and uRLLC superclasses. We aim to adapt the existing model to this new data while retaining knowledge from the initial dataset, updating the model parameters to $\mathcal{\theta}_\tau$, and minimizing the cross-entropy loss over both datasets. The Fig. \ref{cl} depicts the process of continual learning, where the model adapts to a continuous stream of new subclass service data without forgetting previously learned information.
	
	\textbf{Model Training with Initial Data}
	The initial model is trained by minimizing cross-entropy loss using SGD on dataset $\mathcal{D}_0$:
	\begin{equation}
		\mathcal{\theta}_0 = \arg\min_{\boldsymbol{\theta}} \frac{1}{\mathcal{N}_0} \sum_{i=1}^{\mathcal{N}_0} \mathcal{L}_{CE}(f(\mathcal{X}_0(i), \boldsymbol{\theta}_0), \mathbb{S}_0(i)).
		\label{train_init}
	\end{equation}
	
	\textbf{Arrival of New Data and Classes}
	At task $t$, new data may be introduced either under existing subclasses or as completely new subclasses. The model is then trained on this new data while preserving knowledge of previous patterns. We define $\mathcal{D}_{\textrm{new}}^{\textrm{data}}$ for new data within previous subclasses, and $\mathcal{D}_{\textrm{new}}^{\textrm{class}}$ for new subclasses.
	
	\textbf{Model Training with New Data}
	Upon receiving new data $\mathcal{D}_\tau$, the model is retrained to adapt to this new information:
	\begin{equation}
		\mathcal{\theta}_\tau = \arg\min_{\boldsymbol{\theta}_\tau} \frac{1}{\mathcal{N}_\tau} \sum_{i=1}^{\mathcal{N}_\tau} \mathcal{L}_{CE}(f(\mathcal{X}_\tau(i), \boldsymbol{\theta}_\tau), \mathcal{Y}_\tau(i)).
		\label{train_new}
	\end{equation}
	
	\textbf{Average Forgetting (AF) and Performance (AP)}
	AF measures the performance loss on the initial dataset after learning new data, and AP measures performance on the new data.
	AF and AP are two key metrics used to evaluate a model’s ability to retain previously learned knowledge while acquiring new information, which is a common requirement in continual learning and adaptive systems. AF quantifies the degradation in performance on the initial dataset after the model has been trained on new data. It reflects the extent to which the model forgets earlier knowledge due to the influence of subsequent learning phases. AP measures the model’s performance on the newly introduced data, using the updated model parameters. Mathematically, they are defined as:
	\begin{equation}
		\small
		\textrm{AF} = \frac{1}{\mathcal{N}_0} \sum_{i=1}^{\mathcal{N}_0} \left( \mathcal{L}_{CE}(f(\mathcal{X}_0(i), \boldsymbol{\theta}_0)) - \mathcal{L}_{CE}(f(\mathcal{X}_0(i), \boldsymbol{\theta}_\tau)) \right),
		\label{AF}
	\end{equation}
	\begin{equation}
		\textrm{AP} = \frac{1}{\mathcal{N}_0} \sum_{i=1}^{\mathcal{N}_0} \mathcal{L}_{CE}(f(\mathcal{X}_0(i), \boldsymbol{\theta}_\tau)).
		\label{AP}
	\end{equation}
	
	A higher AF indicates greater forgetting of the original knowledge. Together, AF and AP provide a quantitative framework to evaluate learning stability (forgetting) and plasticity (performance). The objective is to minimize average forgetting while maximizing average performance, measured in terms of accuracy. $\theta_{\mathcal{T}}$ is the set of all model parameters, including $\theta_{0}$ and $\theta_{\tau}$. The problem definition aims to find an optimal $\theta_{\textrm{optimal}}$ model that balances forgetting and performance while satisfying the performance threshold.
	\begin{equation}
		\text{CL-S-SP} : \hspace{4mm} \min_{\theta_\mathcal{T}} \hspace{4mm} (\textrm{AF} +   (1 - \mathcal{\textrm{AP}})).
		\label{cl-l-sp}
	\end{equation}
	
	\section{Solution Approach} \label{SA}
\begin{figure}[t]
		\centerline{\includegraphics[width=.95\linewidth]{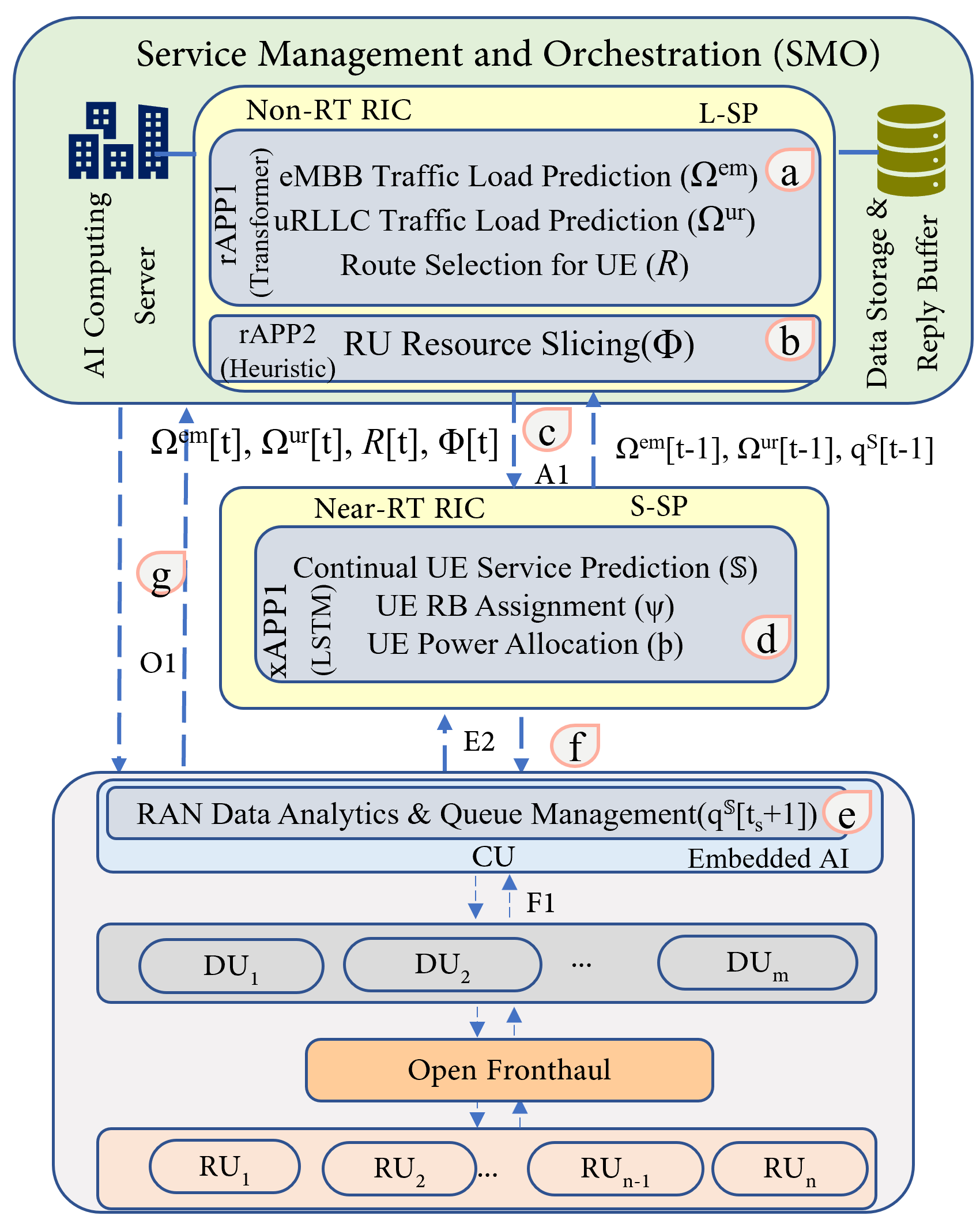}}
		\caption{Proposed intelligent framework for managing multi-service-modal UEs (MSMUs) within the AI-RAN ecosystem. The L-SP is handled in the non-RT RIC through two rApps: rApp1 performs Transformer-based demand and route prediction, while rApp2 applies heuristic-based radio resource slicing. The S-SP is addressed in the near-RT RIC via xApp1, which uses a continual learning-enhanced LSTM model. Queue management decisions are executed at the CU, leveraging the embedded AI/ML capabilities of the AI-RAN architecture.}
		
		\label{solution}
	\end{figure}

	We address the L-SP and the S-SP on distinct time scales. The decomposition of problem (\ref{p1_ran}) into subproblems (\ref{l-sp}) and (\ref{s-sp}) yields a suboptimal solution. However, the obtained solution remains near-optimal, ensuring a minimal optimality gap. Optimal RAN resource slicing $\Phi$ critically depends on accurate predictions of the traffic demand vector $\boldsymbol{\Omega}^\text{em}$ and $\boldsymbol{\Omega}^\text{ur}$, routes vector $\boldsymbol{R}$. To optimize eMBB throughput and minimize uRLLC latency, accurate estimation of traffic data is crucial. However, due to the dynamic network environment and periodic updates from RAN components, obtaining real-time information is challenging. To address this, we propose using historical system data from past time frames via the O1 interface to generate more accurate optimal responses. Our approach incorporates continual learning, allowing the system to adapt to evolving traffic and resource demands, ensuring consistent performance in dynamic environments. The scenarios for high-level deployment are shown in Fig. \ref{solution}. We provide the whole workflow of the proposed algorithm's below, sequentially labeled alphabetically within the AI-RAN architecture.
	
	\begin{enumerate}[label=(\alph*)]
		\item At the beginning of time-frame $t > 1$, the deep learning procedure for solving L-SP is carried out at Non-RT RIC based on the collected RAN data in SMO. For $t = 1$, the traffic demand $\boldsymbol{\Omega}^\text{em}$ and $\boldsymbol{\Omega}^\text{ur}$, and suitable route $\boldsymbol{R}$ for all UEs are initialized by rAPP1(Transformer model).
		
		\item  Based on the predicted traffic demand $\boldsymbol{\Omega}^\text{em}$ and $\boldsymbol{\Omega}^\text{ur}$, suitable route $\boldsymbol{R}$, the bandwidth of each RU  for eMBB and uRLLC are separated by rAPP2.
		
		\item The traffic demand $\boldsymbol{\Omega}^\text{em}[t]$ and $\boldsymbol{\Omega}^\text{ur}[t]$, optimal flow-split decisions (suitable RU $\boldsymbol{R}[t]$), and slicing variable $\Phi[t]$ are sent to Near-RT RIC via A1 (the standardized open interface) for real deployment.
		
		\item Given $\boldsymbol{\Omega}^\text{em}[t]$, $\boldsymbol{\Omega}^\text{ur}[t]$, $\boldsymbol{R}[t]$, $\Phi[t]$, in the near-RT RIC, xAPP1 manages congestion by solving the S-SP and optimizing RAN resources and operations in each time-slot $t_s$. It predicts the service $\mathbb{S}[t_s]$, optimal resource blocks $\boldsymbol{\Psi}[t_s]$ and power vector $\boldsymbol{\mathcal{P}}[t_s]$ for the service.
		
		\item Subsequently, the RAN data analytic component in CU for AI-RAN updates queue-lengths $\boldsymbol{q}^\mathbb{S}[t_{s+1}]$ as in Step 8 of Algorithm 1. The updated queue-lengths are sent back to SMO through the O1 interface for periodic reporting.
		
		\item The xApps hosted in the near-RT RIC communicate with CU/DU through the E2 interface. 
		
		
		\item The performance and inspections (e.g., $\boldsymbol{q}^\mathbb{S}[t-1]$, $\boldsymbol{\Omega}^\text{em}[t-1]$, $\boldsymbol{\Omega}^\text{ur}[t-1]$, $\boldsymbol{R}[t-1]$....) are provided to the SMO over the O1 interface following $S_i$ TTI in order to re-forecast the traffic demand $\boldsymbol{\Omega}^\text{em}_*[t+1]$, $\boldsymbol{\Omega}^\text{ur}_*[t+1]$, route $\boldsymbol{R}^*[t+1]$ and others.
	\end{enumerate}
	
	\begin{algorithm}[t]
		\caption{Proposed MSMUs supported intelligent AI-RAN management}
		\label{alg:ran_management}
		\textbf{Initialization:} Set $t = 1$, $t_s = 1$, $R_u[1] = \frac{\mathbf{1}}{E}[1, \dots, 1]$, and $\Phi[1] = \frac{1}{2}$; every initial queue is configured to be empty, $q^\mathbb{S}_{e,u}[1] = 0$ and $\boldsymbol{q}^\mathbb{S}[1] = 0$. 
		\begin{algorithmic}[1]
			\FOR{$t = 1, 2, \dots, T$}
			\STATE \textbf{Traffic demand and route prediction:} Given $(\boldsymbol{\Omega}^\text{em}[t-1], \boldsymbol{\Omega}^\text{ur}[t-1], \boldsymbol{R}[t-1], \boldsymbol{q}^\mathbb{S}[t-1])$, the rAPP1 predict traffic demand, route and power for all users solving L-SP and rApp2 splits the available bandwidth of all RUs ($\Phi[t]$) users using (\ref{ran_slice}).
			\FOR{$t_s = 1, 2, \dots, S_i$ with $s \in \{1, 2, \dots, S_i\}$}
			\STATE \textbf{Service Identification and Resource Optimization:} Given both services' queue-length vectors $\boldsymbol{q}^\mathbb{S}[t_s]$, and every long-term factor, including $(\boldsymbol{\Omega}^\text{em}_*[t], \boldsymbol{\Omega}^\text{ur}_*[t], \boldsymbol{\Phi}^*[t], \boldsymbol{R}^*[t])$, xAPP1 solve the S-SP to predict the service type $\mathbb{S}^*$, get the RB allocation ($\boldsymbol{\Psi}^*$) and power determination $\boldsymbol{\mathcal{P}}^*$ for the predicted service.
			\STATE \textbf{Updating Queue-Lengths:} AI-RAN data analytic component in CU updates queue-lengths as:
			\begin{align}
				q^\mathbb{S}_{e,u}[t_s + 1] = \max \big\{ & q^\mathbb{S}_{e,u}[t_s] + R_{e,u}[t] \Omega^\mathbb{S}_u[t] Z^\mathbb{S} \delta_i \nonumber \\
				& - \mathcal{R}^\mathbb{S}_{e,u}[t_s] \delta_i, 0 \big\}  \nonumber.
			\end{align}
			\STATE Set $s = s + 1$.
			\ENDFOR
			\STATE Update $\{\boldsymbol{q}^\mathbb{S}[t], \boldsymbol{\Omega}^\mathbb{S}[t]\} = \{q^\mathbb{S}_{e,u}[t], \Omega^\mathbb{S}_u[t]\}, \forall u \in U, e \in \mathcal{E}_f$.
			\STATE Set $t = t + 1$.
			\ENDFOR
		\end{algorithmic}
	\end{algorithm}

	\subsection{Solution Approach for L-SP}
	As mentioned, RAN resource slicing is challenging to optimize due to the uncertainty of key variables such as $\boldsymbol{\Omega}^\text{em}[t]$, $\boldsymbol{\Omega}^\text{ur}[t]$, and $\boldsymbol{R}[t]$ at the start of a timeframe. To take effective policies we require accurate predictions of traffic demand and route. To address this, we adopt deep learning approach. We propose a ReVIN-enhanced Multi-UE Transformer model for the concurrent prediction of traffic demand and routing decisions of multiple UEs simultaneously. Each UE's data sequence is first linearly projected to a hidden representation and then augmented with positional encoding to retain temporal information. These sequences are subsequently processed in parallel by a shared Transformer Encoder. The core of this encoder is a self-attention mechanism, which captures long-range temporal dependencies across all UEs. Although Transformers are highly effective at capturing the complex, bursty patterns characteristic of dynamic RAN environments, their performance is notably sensitive to non-stationary time series, which is the inherent nature of dynamic traffic demands. To mitigate this vulnerability, we integrate ReVIN (Reversible Instance Normalization) which dynamically normalizes the input data to the Transformer, thereby enhancing stability, while preserving the instance-specific statistics that are restored during the decoding phase. We details the step-by-step mechanism for predicting demand and routing using a Transformer model below. 
	
	\subsubsection{Input Data Representation}
	Each UE $u$ provides a sequence of input features over the previous $T$ time frames, represented as:
	\begin{equation}
		\boldsymbol{X}_u = \left[ \boldsymbol{X}_{T}^u, \boldsymbol{X}_{T-1}^u, \dots, \boldsymbol{X}_1^u \right] \in \mathbb{R}^{T \times F},
		\label{idrl}
	\end{equation}
	where each $\boldsymbol{X}_t^u \in \mathbb{R}^F$ is a feature vector for UE $u$ at time step $t$, containing information such as traffic demand ($\boldsymbol{\Omega}^\text{em}$ and $\boldsymbol{\Omega}^\text{ur}$) and routing decisions ($\boldsymbol{R}$) for the previous $T$ time steps. For all $U$ UEs, the input matrix is represented as:
	\begin{equation}
		\boldsymbol{X} = { \boldsymbol{X}_1, \boldsymbol{X}_2, \dots, \boldsymbol{X}_U } \in \mathbb{R}^{U \times T \times F},
		\label{idral}
	\end{equation}
	where $\boldsymbol{F} = {\boldsymbol{\Omega}^\text{em}, \boldsymbol{\Omega}^\text{ur}, \boldsymbol{R} }$. To address the non-stationarity and varying distribution of input features across users and time, we employ ReVIN~\cite{kim2022reversible} as a preprocessing step to normalize each UE's temporal data before it is fed to the Transformer encoder, preserving instance-specific statistics for accurate decoding reversion.

	\subsubsection{Reversible Instance Normalization (ReVIN)}
	ReVIN normalizes the input per instance and allows the model to learn in a stable and normalized space while preserving the ability to revert predictions back to the original scale. Let $\boldsymbol{X} \in \mathbb{R}^{B \times T \times U \times F}$ be a batch of input sequences, where $B$ is batch size.	In normalization mode, ReVIN computes the mean and standard deviation of the input $\boldsymbol{X}$ across the batch, time, and user dimensions:
	\begin{equation}
		\mu_\text{rev} = \mathbb{E}_{b,t,u}[\boldsymbol{X}_{b,t,u,:}], \quad 
		\sigma_\text{rev} = \sqrt{\mathbb{V}_{b,t,u}[\boldsymbol{X}_{b,t,u,:}]} + \varepsilon,
	\end{equation}
	where $\mu_\text{rev}, \sigma_\text{rev} \in \mathbb{R}^{1 \times 1 \times 1 \times F}$ are the per-feature mean and standard deviation, and $\varepsilon$ is a small constant added for numerical stability. If affine transformation is enabled, ReVIN learns per-feature scale and shift parameters $\gamma_\text{rev}, \beta_\text{rev} \in \mathbb{R}^{1 \times 1 \times 1 \times F}$. The normalized input with affine transformation is computed as:
	\begin{equation}
		\widetilde{\boldsymbol{X}} = \gamma_\text{rev} \cdot \left( \frac{\boldsymbol{X} - \mu_\text{rev}}{\sigma_\text{rev}} \right) + \beta_\text{rev}.
	\end{equation}
    If affine transformation is disabled, $\gamma_\text{rev}$ and $\beta_\text{rev}$ default to 1 and 0, respectively. During inference or output post-processing, ReVIN reverses the normalization using the stored $\mu_\text{rev}$ and $\sigma_\text{rev}$ values. If affine transformation was used, it is first reversed then denormalization:
	\begin{equation}
		\widehat{\boldsymbol{Y}} = \left( \boldsymbol{Y} - \beta_\text{rev} \right) / (\gamma_\text{rev} + \varepsilon),
	\end{equation}
	\begin{equation}
		\boldsymbol{Y}_{\text{denorm}} = \widehat{\boldsymbol{Y}} \cdot \sigma_\text{rev} + \mu_\text{rev}.
	\end{equation}
	By this the model's predictions are mapped back to the original feature scale, making them directly interpretable (e.g., for traffic demand in Mbps).
	\begin{figure}[!t]
		\centerline{\includegraphics[width=\linewidth]{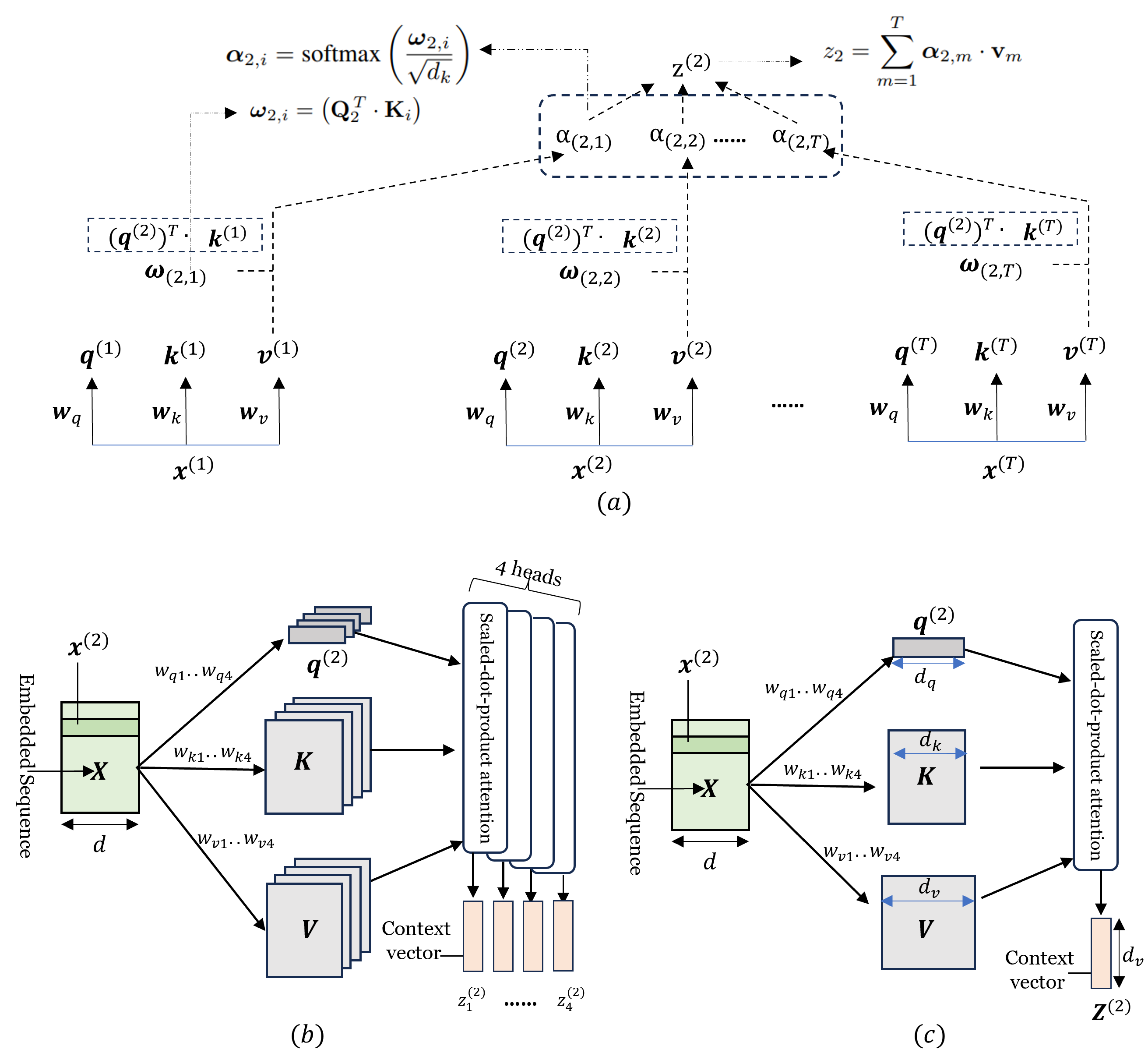}}
		\caption{(a) Overview of the process for computing single-head attention scores; (b) visualization of the single-head attention mechanism; (c) illustration of the multi-head attention architecture, adapted from \cite{advancing}.}
		\label{transformer}
	\end{figure}
	
	\subsubsection{Transformer Processing and Output Generation}
	Unlike sequential models such as LSTM/GRU, the Transformer leverages self-attention mechanisms to process entire sequences in parallel, enabling it to efficiently capture both short- and long-range dependencies in dynamic RAN environments. Before feeding the input into the Transformer, we apply ReVIN to the feature sequence of each UE. This operation standardizes the input statistics while preserving the ability to later revert to the original scale. The normalized input for UE $u$ is denoted as $\widetilde{\boldsymbol{X}}_u \in \mathbb{R}^{T \times F}$.
	
	\paragraph{Scaled Dot-Product Self-Attention}  
	At the heart of the Transformer lies the self-attention mechanism, which enables the model to weigh the relevance of different time steps when generating contextual representations. Each time step in the normalized input is projected into three distinct spaces to compute queries ($\boldsymbol{Q}$), keys ($\boldsymbol{K}$), and values ($\boldsymbol{V}$), using learned weight matrices:
	\begin{equation}
		\boldsymbol{Q} = \widetilde{\boldsymbol{X}}_u \boldsymbol{W}^Q,\quad
		\boldsymbol{K} = \widetilde{\boldsymbol{X}}_u \boldsymbol{W}^K,\quad
		\boldsymbol{V} = \widetilde{\boldsymbol{X}}_u \boldsymbol{W}^V,
	\end{equation}
	where $\boldsymbol{W}^Q, \boldsymbol{W}^K, \boldsymbol{W}^V \in \mathbb{R}^{F \times d_k}$ are learnable parameters, and $d_k$ is the dimension of the key/query vectors. The attention scores are computed as the scaled dot product between queries and keys, followed by a softmax operation to obtain attention weights:
	\begin{equation}
		\text{Attention}(\boldsymbol{Q}, \boldsymbol{K}, \boldsymbol{V}) = \text{softmax}\left( \frac{\boldsymbol{Q} \boldsymbol{K}^\top}{\sqrt{d_k}} \right) \boldsymbol{V}.
	\end{equation}
	This mechanism allows each time step to attend to others, assigning higher weights to more relevant positions. Fig. \ref{transformer}(a) illustrates the computation of a context vector, $z$, for an input sequence $X = \{x_1, x_2, \ldots, x_T\}$ of length $T$, using the self-attention mechanism. Fig. \ref{transformer}(b) illustrate a single attention head to process the input sequence using the scaled dot-product attention mechanism. First, the input sequence is linearly projected into three distinct vector spaces by the weight matrices $W^q$, $W^k$, and $W^v$, creating sequences of Queries ($Q$), Keys ($K$), and Values ($V$). These projections are tailored to dimensions $d_k$, $d_k$, and $d_v$ respectively. The raw attention weights, $\omega$, are then computed by taking the dot product of the Queries and Keys, scaled by a factor of $1/\sqrt{d_k}$. Subsequently, these weights are normalized using a softmax function to produce the attention scores, $\alpha$. Finally, the context vector, $z$, is generated by taking a weighted sum of the Value vectors, where the attention scores, $\alpha$, serve as the dynamic weights. This process allows the model to selectively focus on the most relevant parts of the input sequence. The multi head attention mechanism is illustrated in Fig. \ref{transformer}(c).
	
	\paragraph{Multi-Head Attention}  
	To capture different types of temporal dependencies simultaneously, we employ multi-head attention, which consists of $H$ parallel attention heads. Each head independently computes self-attention using separate projection matrices, capturing diverse perspectives on the input:
	\begin{equation}
		\hspace{-3mm}
		\text{MultiHead}(\boldsymbol{Q}, \boldsymbol{K}, \boldsymbol{V}) = \text{Concat}(\text{head}_1, \dots, \text{head}_H) \boldsymbol{W}^O,
	\end{equation}
	where each attention head is defined as $\text{head}_i = \text{Attention}(\boldsymbol{Q}_i, \boldsymbol{K}_i, \boldsymbol{V}_i)$ and $\boldsymbol{W}^O$ is a learnable output projection matrix.
	
	\paragraph{Feed-Forward Network (FFN)}  
	The output of the multi-head attention layer is then passed through a position-wise feed-forward network. This FFN consists of two fully connected layers with a ReLU activation in between, applied independently at each time step:
	\begin{equation}
		\text{FFN}(\boldsymbol{z}) = \text{ReLU}(\boldsymbol{z} \boldsymbol{W}_1 + \boldsymbol{b}_1) \boldsymbol{W}_2 + \boldsymbol{b}_2,
	\end{equation}
	where $\boldsymbol{W}_1$, $\boldsymbol{W}_2$, $\boldsymbol{b}_1$, and $\boldsymbol{b}_2$ are learnable parameters. This stage enriches the feature representations with non-linear transformations.
	
	\paragraph{Sequence Aggregation and Output Generation}  
	After self-attention and feed-forward operations within the Transformer encoder, the temporal sequence is mapped into a compact context vector $\boldsymbol{H}_u$ for each UE $u$. Rather than using global average pooling, we extract the representation from the final time step, which captures accumulated temporal information:
	\begin{equation}
		\boldsymbol{H}_u = \text{TransformerEncoder}(\text{ReVIN}(\boldsymbol{X}_u))_{T} \in \mathbb{R}^{d}.
	\end{equation}
	This output embedding $\boldsymbol{H}_u$ is then forwarded to multiple task-specific fully connected heads to predict various RAN metrics:
	\begin{equation}
		\boldsymbol{y}_u = \left[ \boldsymbol{y}_{\Omega^\text{em}}^u, \boldsymbol{y}_{\Omega^\text{ur}}^u, \boldsymbol{y}_{R}^u \right],
	\end{equation}
	where $\boldsymbol{y}_{\Omega^\text{em}}^u \in \mathbb{R}$ is the predicted eMBB traffic demand (continuous), $\boldsymbol{y}_{\Omega^\text{ur}}^u \in \mathbb{R}$ is the predicted uRLLC traffic demand (continuous), $\boldsymbol{y}_{R}^u \in \{1,2,3,4\}$ is the predicted route. The detailed architecture or our proposed transformer is outlined in Table~\ref{tab:multiue_transformer}.
	\begin{table}[!t]
		\centering
		\caption{Architecture of our proposed Multi-UE Transformer Model}
		\label{tab:multiue_transformer}
		\begin{tabular}{@{}ll@{}}
			\text{Component} & \textbf{Description} \\ \hline
			\text{Input} & Tensor shape: $(\text{batch}, \text{no\_of\_UEs}, \text{seq\_len}, \text{features})$ \\
			\text{Reshape Layer} & Reshape to $(\text{batch} \times \text{no\_of\_UEs}, \text{seq\_len}, \text{features})$ \\
			\text{Linear Projection} & Linear layer: $\text{Linear}(\text{features}, 64)$ \\
			\text{Positional Encoding} & Add sinusoidal encoding to input embeddings \\
			\text{Dropout} & Dropout with $p = 0.3$ applied after encoding \\
			\multirow{2}{*}{\text{Transformer Encoder}} & 6-layer encoder with 8 heads and FFN dim = 512,\\ & GELU activation \\
			\text{Permutation} & Input permuted to $(\text{seq\_len}, \text{batch}, 64)$ for Transf. \\
			\text{Temporal Aggregation} & Extract output from last time step \\
			\text{Dropout Layer} & $0.3$ applied to final time-step embedding \\
			\text{Reshape Back} & Reshape to $(\text{batch}, \text{no\_of\_UEs}, 64)$ \\
			\text{Output Heads} &
			\begin{tabular}[c]{@{}l@{}}
				$-$ eMBB Demand: $\text{Linear}(64, 1)$ -MSE Loss\\
				$-$ uRLLC Demand: $\text{Linear}(64, 1)$ -MSE Loss\\
				$-$ Route: $\text{Linear}(64, \text{no\_of\_RU})$ -Cross Entropy
			\end{tabular} \\ \hline
		\end{tabular}
	\end{table}
	
	\paragraph{ReVIN Denormalization}  
	As ReVIN normalization was applied, the predicted outputs must be denormalized to restore them to their original scale. This is especially important for eMBB and uRLLC demand, which involve continuous (regressive) values. For categorical predictions, such as route classification, denormalization is not required. However, for regression-based outputs like demand forecasting, denormalization is essential to interpret the results accurately. We de-normalized the demand to restore the original data scale as follows:
    \begin{equation}
    		\widehat{\boldsymbol{y}}_u = \text{ReVIN}^{-1}(\boldsymbol{y}_u, \mu_\text{rev}, \sigma_\text{rev}),
    \end{equation}
	where $\mu_\text{rev}$ and $\sigma_\text{rev}$ are the statistics computed during the normalization step.
		
	Long-term data collected from the RAN environment is utilized to train the Transformer-based model within the AI-RAN architecture, specifically at the non-RT RIC. Algorithm~\ref{alg:Transformer_RAN_Slicing} presents the operational flow of the proposed Transformer-driven framework for long-term metric forecasting in RAN slicing. Once trained, the model is deployed to the non-RT RIC and connected to the near-RT RIC via the A1 interface. The Transformer's predictions guide intelligent RAN management, ensuring adaptive slicing in response to dynamic traffic patterns and service requirements.

		\begin{algorithm}[t]
			\caption{Transformer-based framework for solving the L-SP problem by predicting eMBB and uRLLC demands, forecasting routes, and optimizing RAN resource slicing.}
			\label{alg:Transformer_RAN_Slicing}
			\begin{algorithmic}[1]
				\REQUIRE Sequence of input features $\boldsymbol{X}_u$ for each UE $u$ over $T$ time frames
				\ENSURE Predicted metrics for each UE $u$: eMBB traffic demand ($\boldsymbol{y}_{\Omega^\text{em}}^u$), uRLLC traffic demand ($\boldsymbol{y}_{\Omega^\text{ur}}^u$), service type ($\boldsymbol{y}_\mathbb{S}^u$), and route class ($\boldsymbol{y}_{R}^u$)
				
				\IF{\text{(Training == True)}}
				\STATE Prepare input using (\ref{idrl}) and (\ref{idral}) for each UE $u$
				\STATE Apply ReVIN normalization to input sequences: $\widetilde{\boldsymbol{X}}_u = \text{ReVIN}(\boldsymbol{X}_u)$
				\STATE Encode sequence using Transformer encoder and extract hidden vector $\boldsymbol{H}_u$ via final time step
				\STATE Predict targets using fully connected output heads: $\boldsymbol{y}_u = \left[ \boldsymbol{y}_{\Omega^\text{em}}^u, \boldsymbol{y}_{\Omega^\text{ur}}^u, \boldsymbol{y}_{R}^u \right]$
				\STATE Compute MSE loss for continuous outputs ($\boldsymbol{y}_{\Omega^\text{em}}^u, \boldsymbol{y}_{\Omega^\text{ur}}^u$) and CrossEntropy for classification output ($\boldsymbol{y}_{R}^u$)
				\STATE Update model parameters using Adam optimizer
				\STATE Save trained model weights
				\ENDIF
				
				\IF{\text{(Inference == True)}}
				\FOR{each time step $t$}
				\FOR{each UE $u$}
				\STATE Construct input feature sequence $\boldsymbol{X}_u$ at time $t$ using (\ref{idrl})
				\STATE Normalize using ReVIN: $\widetilde{\boldsymbol{X}}_u = \text{ReVIN}(\boldsymbol{X}_u)$
				\STATE Encode using Transformer: $\boldsymbol{H}_u = \text{TransformerEncoder}(\widetilde{\boldsymbol{X}}_u)_T$
				\STATE Generate predictions using output heads
				\STATE Apply ReVIN denormalization for continuous outputs ($\boldsymbol{y}_{\Omega^\text{em}}^u, \boldsymbol{y}_{\Omega^\text{ur}}^u$)
				\STATE Perform RAN slicing by (\ref{ran_slice})
				\ENDFOR
				\ENDFOR
				\ENDIF
			\end{algorithmic}
		\end{algorithm}
		
		\subsubsection{Radio Resource Slicing}
		We posit that the queue length of data flows $u$ in the upcoming frame depends on the traffic demand, as well as on the RU selection of the flow from $u$ in both the running and recent frames. We assume the network parameter, denoted as $\boldsymbol{\Omega}^\text{em}_u$, $\boldsymbol{\Omega}^\text{ur}_u$, $\boldsymbol{R}$ collectively represents the dynamic data arrival rate or eMBB and uRLLC across all cells of RUs and are transmitted immediately to the rAPP2 for enhancing the dynamic bandwidth split in non-RT RIC, $\Phi[t]$. These parameters are developed on a longer time scale, i.e., on the full frame basis rather than the mini timeslot basis of resource block assignment and power allocation, for efficient deployment. It is necessary to ascertain $\Phi[t]$ at the start of every frame. It is particularly challenging to determine the ideal bandwidth split and flow split values since the CSI of upcoming slots in the present frame is uncertain. In order to ascertain $\Phi[t]$, we therefore suggest an effective heuristic algorithm based on $\boldsymbol{\Omega}^{\text{em},*}[t]$, $\boldsymbol{\Omega}^{\text{ur},*}[t]$, and $\boldsymbol{R}^*[t]$. Allocating bandwidth to each service in proportion to the associated traffic demands for each RU is a logical approach. However, this approach is ineffective at satisfying the strict latency requirements of uRLLC applications since the volume of uRLLC traffic is significantly lower than that of eMBB traffic. We address this by taking into account the overall traffic demands as well as the maximum acceptable delays for both services. Consequently, the following formula is used to determine the bandwidth difference between eMBB and URLLC services:
		
		\begin{equation}
			\Phi_e^*[t] = \frac{\sum_{\mathbb{S} \in \mathbb{S}_{\text{ur}}} \Omega^{\mathbb{S},*}_{u,e}[t]}{\sum_{\mathbb{S} \in \mathbb{S}_{\text{em}}} \Omega^{\mathbb{S},*}_{u,e}[t]} \times \frac{\Upsilon_{\text{em}}^{\text{th}}}{\Upsilon_{\text{ur}}^{\text{th}}},
			\label{ran_slice}
		\end{equation}
		where $\Upsilon_{\text{ur}}^{\text{th}}$ and $\Upsilon_{\text{em}}^{\text{th}}$ represent the uRLLC and eMBB services' respective maximum permitted latency.
		
		\subsection{Solution Approach for S-SP}
		After performing RAN resource slicing, all relevant values are forwarded to the near-RT RIC to address the S-SP. Although power ($\boldsymbol{\mathcal{P}}[t_s]$) and PRBs ($\boldsymbol{\Psi}[t_s]$) can be optimized directly, the S-SP remains challenging for optimization techniques. This is because key variables, such as service type ($\mathbb{S}[t_s]$), are often uncertain at the beginning of a mini time frame. Effective policies require accurate predictions of service for next mini time frame. To address the issue we adopt meta-heuristic approach for solving the S-SP. We need lightweight model that can be operated smoothly in real time fulfilling near-RT-RIC constraint requirement, therefore, we propose lightweight Multi-UE LSTM Model for predicting service type for that mini timeframe and suitable PRBs and power for that service. The LSTM model is designed for $U$ UEs and processes input data from the current timeframe, received from the non-RT RIC, along with the outcomes from the previous mini timeframe within the current timeframe. Each LSTM layer receives time-series input of shape $(\text{seq\_len}, \text{input\_size})$ and produces hidden representations at each time step. The hidden state corresponding to the final time step is extracted, regularized using dropout, and passed through three fully connected layer with ReLU activation to ensure service type prediction, non-negative resource block predictions, and power.
		
		\begin{table}[t]
			\centering
			\caption{Architecture of the proposed Multi-UE LSTM Model}
			\label{tab:multiue_lstm_architecture}
			\begin{tabular}{@{}ll@{}}
				\text{Component} & \textbf{Description} \\ \hline
				\text{Input} & Tensor shape: $(\text{batch}, \text{no\_of\_UE}, \text{seq\_len}, \text{feature})$ \\
				\multirow{5}{*}{\text{UE-specific LSTMs} }& \text{no\_of\_UE} independent LSTM layers: \\
				& $\-$ Input size = \text{feature} \\
				& $-$ Hidden size = 32 \\
				& $-$ Number of layers = 2 \\
				& $-$ Batch-first = True \\
				\text{Last Time Step} & Extract final hidden state per UE: $(\text{batch}, 32)$ \\
				\text{Dropout} & Dropout with $p = 0.2$ on LSTM output \\
				\text{Stack Outputs} & Stack across UEs to shape $(\text{batch}, \text{no\_of\_UE}, 32)$ \\
				\text{Activation Function} & ReLU to ensure non-negative predictions \\
				\text{Output} & 	\begin{tabular}[c]{@{}l@{}}
					$-$ Service: $\text{Linear}(64, 2)$ -Cross Entropy\\
					$-$ PRBs: $\text{Linear}(64, 1)$ -MSE Loss\\
					$-$ Power: $\text{Linear}(64, 1)$ - MSE Loss \\ 
				\end{tabular}\\ \hline
			\end{tabular}
		\end{table}
		
		\subsubsection{Input Data Representation}
		The input feature for each UE $ u $ at mini time step $ t_s $ is represented by a feature vector: $\boldsymbol{X}_{t_s}^{u} \in \mathbb{R}^{F+f}$. Where, $ F $ is the number of features from L-SP ($\boldsymbol{\Omega}^\text{em}$, $\boldsymbol{\Omega}^\text{ur}$, $\boldsymbol{R}$, and $\boldsymbol{\Phi}$) from the non RT RIC. $f$ is the feature of mini time frame (service, PRBs, and power). The input data for all $ U $ UEs over previous $T$ mini steps is represented by a 3-dimensional tensor:
		\begin{align}
			&\{\boldsymbol{X}_{t_s}^{u} \in \mathbb{R}^{F+f} \mid u = 1, 2, \dots, U; T = 1, 2, \dots, t_s \}, 
			\label{idr1}
		\end{align}
		\begin{align}
			&\boldsymbol{X} \in \mathbb{R}^{U \times (F + T \cdot f)},
			\label{idr2}
		\end{align}
		where, $\boldsymbol{F} = \{\boldsymbol{\Omega}^\text{em}, \boldsymbol{\Omega}^\text{ur}, \boldsymbol{R}, \boldsymbol{\Phi}\}$ from the current time step and $f = \{\mathbb{S}, \boldsymbol{\Psi}, \boldsymbol{\mathcal{P}}\}$. This input tensor has dimensions $\boldsymbol{X} \in \mathbb{R}^{N \times (F+T.f)}$. Here the $\text{sequence number} = \text{maximum mini frame} -1$ under a time frame. For the first mini frame we will make the all previous sequence 0, for second we will keep the last one and make the rest 0 in such way for the last mini slot we will keep the previous value for all mini slot.
		\begin{algorithm}[t]
			\caption{Algorithm for solving the S-SP problem through service type, PRBs, and optimal power prediction.}
			\label{alg:GRU_PRB_Prediction}
			\begin{algorithmic}[1]
				\STATE \textbf{Input:} Data tensor $\boldsymbol{X} = \{\boldsymbol{X}_{t_s}^u \in \mathbb{R}^{F+f} \mid u = 1, 2, \dots, U; T = 1, 2, \dots, t_s\}$ for $U$ UEs, $F$ for that time frame and $f$ for over all previous mini time frames for that time frame
				\STATE \textbf{Output:} Predicted service type ($\boldsymbol{Y}_\mathbb{S}$), PRB vectors ($\mathbf{\boldsymbol{Y}_\Psi}$), and power vectors ($\boldsymbol{Y}_\mathcal{P}$) for all UEs $U$.
				\IF{\text{(Training == True)}}
				\STATE Prepare input data using (\ref{idr1}) and (\ref{idr2})
				\STATE Evaluate Constraints: (\ref{s-sp:const1})-(\ref{s-sp:const9})
				\STATE Calculate CrossEntropy Loss for $\mathbb{S}$, and MSE loss for $\Psi$ and $\mathcal{P}$ then optimize the LSTM model using Adam Optimizer
				\STATE Save Trained Model Parameters
				\ENDIF
				\IF{\text{(Inference == True)}}
				\FOR{each time step $t_s$}
				\FOR{each UE $u$}
				\STATE Define the input feature representation $\boldsymbol{X}^u_{t_s}$ using (\ref{idr1}) containing features at mini timestep $t_s$
				\ENDFOR
				\STATE Make the input feature representation for all $U$ by (\ref{idr2})
				\STATE Input to LSTM cell and execute operation.
				\STATE Obtain the final hidden state $\boldsymbol{H}_{t_s}^u$ for each UE	
				\STATE Predict the service, PRBs, and power for each UE by (\ref{og1}))
				\ENDFOR
				\ENDIF
			\end{algorithmic}
		\end{algorithm}
		
		\subsubsection{LSTM Processing and Outputs Generation}  
		At each mini-time step $ t_s $, the LSTM processes the input $ \boldsymbol{X}_{t_s}^u \in \mathbb{R}^{F + f} $ together with the hidden state from the previous time step $ \boldsymbol{H}_{t_s-1}^u \in \mathbb{R}^H $ and the cell state $ \boldsymbol{C}_{t_s-1}^u \in \mathbb{R}^H $. After processing the sequence through the LSTM layers, the hidden state $ \mathbf{H}_{t_s}^u $ at the last time step $ t_s $ is used for making predictions. The outputs are service type, PRBs, and power for each UE for that time frame. The final hidden state $ \mathbf{H}_{t_s}^u $ is used to predict three different metrics for each UE $ u $. The output for UE $ u $ is denoted as:
		\begin{equation}
			\boldsymbol{Y}_u = \left[ \boldsymbol{Y}^u_\mathbb{S}, \boldsymbol{Y}_\Psi^u, \boldsymbol{Y}_\mathcal{P}^u \right],
			\label{og1}
		\end{equation}
		where,
		$\boldsymbol{Y}_\mathbb{S}^u \in \mathbb{R}^2$ is the predicted service (categorical: eMBB/uRLLC),
		$\boldsymbol{Y}_\Psi^u \in \mathbb{R}$ is the predicted PRBs (continuous), and
		$\boldsymbol{Y}_\mathcal{P}^u \in \mathbb{R}$ is the predicted power (continuous).
		The hidden state $ \mathbf{H}_{t_s}^u $ captures the relevant information from the entire sequence of inputs, allowing the model to make informed predictions about resource allocation. Scaling the PRBs values within 0-1, we determine how many discrete PRBs will be allocated in a UE for that mini time frame. Thus the LSTM captures the temporal dependencies for each UE, and the final output provides a unified prediction for the next mini time frame. We attain a near-optimal solution through the LSTM-based approach, which converges after a few training epochs, closely approximating the optimal solution. The detailed architecture or our proposed LSTM is outlined in Table \ref{tab:multiue_lstm_architecture}. The operational flow of the S-SP framework is presented in Algorithm~\ref{alg:GRU_PRB_Prediction}. This process allows the model to make informed predictions about future mini-time frames, optimizing the allocation of resources based on the input data and learned temporal patterns.
		
		\begin{algorithm}[t]
			\caption{Algorithm for incremental service learning}
			\label{alg_continual_learning}
			\begin{algorithmic}[1]
				\renewcommand{\algorithmicrequire}{\textbf{Input:}}
				\renewcommand{\algorithmicensure}{\textbf{Output:}}
				\REQUIRE Initial model parameters $\theta_\text{init}^\mathbb{S}$.
				\ENSURE Trained model parameters $\theta_\text{new}^\mathbb{S}$.
				\\ \textbf{Initialization:} Memory for task-specific information: $\mathcal{M}_{S} = \mathcal{D}_\text{init}^\mathbb{S}(\mathcal{X}_\text{init}, \mathbb{S}_\text{init})$ for initial task , where $\mathcal{X}_\text{init}$ is the feature matrix, and $\mathcal{Y}_\text{init}$ is the label set for initial task.
				\FOR{each $\mathcal{D}_\mathcal{T}^\mathbb{S}(\mathcal{D}_{nd}^\mathbb{S}, \mathcal{D}_{nc}^\mathbb{S}$) task arrival}
				\FOR{each $\mathcal{D}_\text{new}^\mathbb{S}(\mathcal{X}_\text{new}, \mathbb{S}_\text{new})$ arrival for task $\tau$}
				\STATE Update memory for task $t$ with new data by (\ref{update_mem}).
				\STATE Train the model with updated $\mathcal{M}_\mathbb{S}$ using (\ref{train_new}).
				\ENDFOR
				\STATE \textbf{Compute} the overall average forgetting using (\ref{AF}).
				\STATE \textbf{Calculate} the average performance using (\ref{AP}).
				\STATE \textbf{Optimize} the objective function in (\ref{cl-l-sp})
				\STATE \textbf{Update} the buffer memory
				\ENDFOR    
			\end{algorithmic} 
		\end{algorithm}
		\subsection{Continual Learning for Solving CL-S-SP}
		To address the CL-S-SP defined in (\ref{cl-l-sp}), we propose a reply-exemplar-based continual learning framework. This framework enables the proposed LSTM model, operating as an xApp1 in the near-RT RIC, to learn new traffic types without forgetting previous ones. In our approach, a replay buffer is maintained within the non-RT RIC. This buffer stores a representative set of samples from each task, allowing the model to revisit and retain previously learned patterns. The key steps of our proposed method are outlined below.
		
		\subsubsection{Initialization}
		We define the initial LSTM model parameters as $\theta_{0}$, memory for task-specific information: $\mathcal{M} = \{(\mathcal{X}_i^\tau, \mathbb{S}_i^\tau)\}$ for task $\tau$, where $\mathcal{X}_i^\tau$ is the feature matrix, and $\mathbb{S}_i^\tau$ is the label set for task $\tau$.
		\subsubsection{Model training and parameter updates with new data arrival}
		For new data arrival of task $t$ we define the memory update as follows:
		\begin{equation}
			\mathcal{M} \leftarrow \mathcal{M} \cup \{(\mathcal{X}_\text{new}^\tau, \mathbb{S}_\text{new}^\tau)\}.
			\label{update_mem}
		\end{equation}
		\begin{table}[t]
			\centering
			\caption{Simulation Parameters}
			\label{tab:simulation_parameters}
			\begin{tabular}{p{4.5cm}p{3cm}}
				\hline
				\textbf{Parameter} & \textbf{Value} \\ \hline
				\multicolumn{2}{c}{\textbf{Network Parameters}} \\ \hline
				No. of RUs & 4 \\ 
				Coverage area of RU & $0.11 \, \text{km}^2$ \\ 
				No. of Dual Mode UE & 24 \\ 
				UE Mobility (moving) & 3 m/s \\ 
				Bandwidth of RU & 3 MHz \\ 
				Downlink Frequency & 0.98 GHz \\ 
				Uplink Frequency & 1.02 GHz \\ 
				Length of time-frame & 10 ms \\ \hline
				\multicolumn{2}{c}{\textbf{Model Parameters}} \\ \hline
				Activation & ReLU, SoftMax \\ 
				Optimizer & Adam \\ 
				Learning Rate & 0.0001 \\ 
				No. of epochs & 100 \\ 
				Framework & PyTorch \\ 
				Language & Python \\ \hline
				\multicolumn{2}{c}{\textbf{Traffic Parameters}} \\ \hline
				uRLLC Traffic Type & Poisson \\ 
				eMBB Traffic Type & Constant bitrate \\ \hline
				\multicolumn{2}{c}{\textbf{Traffic Patterns}} \\ \hline
				uRLLC Rate & 10 pkt/s \\ 
				uRLLC Packet Size & 125 bytes \\ 
				eMBB Rate & 1 Mbps \\ \hline
			\end{tabular}
		\end{table}
		
		Then we train the LSTM model with the updated memory $\mathcal{M}$ to obtain task-specific model parameters.
		where $\mathcal{L}$ is the task-specific cross entropy loss function. Then we compare the initial performance got by (\ref{train_init}) with the updated performance got by (\ref{train_new}) after training with new data. We measure the AF and AP by (\ref{AF}), and (\ref{AP}) respectively. Then we minimize the objective function by sufficient training epochs. Repeat this process as new tasks or data ($\tau+2$, $\tau+3$, ...) arrive, updating the memory and training the model accordingly. The algorithm of the proposed incremental service categorization is shown in Algorithm \ref{alg_continual_learning}.
		
		\section{Simulation Results and Analysis} \label{Ex}	
		\subsection{Dataset and Simulation Details}
		We use the Colosseum O-RAN COMMAG dataset \cite{Bonati}, a sophisticated dataset developed for advancing research in 5G and beyond O-RAN networks. The dataset simulates a dense urban setting modeled on data from Rome, Italy, presenting a 5G network scenario involving 4 RUs and 40 user UEs. These UEs navigate and interact within this high-density urban environment, reflecting real-world complexities. We customized the dataset according to our proposed system model by selecting 24 UEs out of the available 40, comprising 12 eMBB and 12 URLLC UEs. Considering the dual-mode capability of UEs, we enabled traffic exchange between eMBB and URLLC UEs sequentially across all RUs. We introduced a new column to classify traffic types (eMBB/uRLLC) and assigned specific RUs to each traffic packet. This setup allows us to effectively manage and analyze traffic flow between different service types in our network model. For the continual service learning we use the ISCXVPN dataset \cite{dataset2}. From the dataset we use 15 classes non vpn application data. We categorized these into 8 applications for eMBB and 7 for URLLC services. The applications were then divided into 8 sequential tasks, each containing one eMBB and one URLLC application, with the final task focused solely on eMBB traffic. This organization allows for systematic and targeted continual learning across different service types. We used an 80:20 train-test split for all tasks to ensure robust model evaluation. We implemented our model using Python 3.10.12 \cite{python}, PyTorch 2.0 \cite{pytorch}. We took 1000 traffic samples in the reply memory from each task containing all five traffic classes of each task. We use Adam optimizer \cite{adam} with a learning rate set to 0.001 and activation function ReLU \cite{relu} for continuous value and Softmax \cite{softmax} for discrete value prediction.  The simulation parameters is shown in Table \ref{tab:simulation_parameters}.
		
		\subsection{Evaluation Metrics}
		We use MSE loss for evaluating demand, power and PRBs prediction and standard accuracy for evaluating the service, and routes prediction. When evaluating our model's effectiveness in continual learning, we utilize task specific accuracy (TSA), task specific forgetting (TSF), average accuracy (AA), average forgetting (AF), catastrophic forgetting (CF), backward transfer (BWT), and forward transfer (FWT) as in \cite{gain_noms}.
		\begin{table}[t]
			\centering
			\renewcommand{\arraystretch}{1.2}
			\caption{MSE Loss between predicted eMBB and uRLLC traffic demand by Transformer, LSTM and GRU compared to the ground truth for both static and moving UEs scenarios.}
			\label{demand_result}
			\begin{tabular}{lcccc}
				\hline
				\textbf{Model} & \multicolumn{2}{c}{\textbf{Static UEs Demand}} & \multicolumn{2}{c}{\textbf{Moving UEs Demand}} \\
				& \textbf{eMBB} & \textbf{uRLLC} & \textbf{eMBB} & \textbf{uRLLC} \\ 
				\hline
				LSTM & 0.00312 & 0.00331 & 0.00362 & 0.00324 \\
				GRU  & 0.00311 & 0.00492 & 0.00435 & 0.00542 \\
				TRANSFORMER  & $\mathbf{0.00295}$ & $\mathbf{0.00302}$ & $\mathbf{0.00313}$ & $\mathbf{0.00288}$ \\
				\hline
			\end{tabular}
		\end{table}
		\begin{table}[!t]
			\centering
			\renewcommand{\arraystretch}{1.2}
			\caption{Accuracy of predicted services and routes by Transformer, LSTM and GRU for both static and moving UEs scenario. }
			\label{accuracy_result}
			\begin{tabular}{lcccc}
				\hline
				\multirow{2}{*}{\textbf{Model}} & \multicolumn{2}{c}{\textbf{Static UEs}} & \multicolumn{2}{c}{\textbf{Moving UEs}} \\
				& \textbf{Service} & \textbf{Route}  & \textbf{Service} & \textbf{Route}  \\ 
				LSTM & $\mathbf{99.86}$ & $\mathbf{99.85}$ & $\mathbf{99.81}$ & 99.81 \\
				GRU & 96.20 & 99.33 & 93.03 & 93.12  \\
				TRANSFORMER  & - & 99.81 & - & $\mathbf{99.85}$ \\
				\hline
			\end{tabular}
		\end{table}
		\begin{table}[!t]
			\centering
			\renewcommand{\arraystretch}{1.2}
			\caption{MSE Loss between predicted PRBs and power by LSTM and GRU compared to the ground truth for both static and moving UEs scenario. }
			\label{resource_result}
			\begin{tabular}{lcccc}
				\hline
				\multirow{2}{*}{\textbf{Model}} & \multicolumn{2}{c}{\textbf{Static UEs}} & \multicolumn{2}{c}{\textbf{Moving UEs}} \\
				& \textbf{PRBs} & \textbf{Power}  & \textbf{PRBs} & \textbf{Power}  \\ 
				LSTM & $\mathbf{0.00302}$ & $\mathbf{0.00245}$ & $\mathbf{0.00304}$ & $\mathbf{0.00294}$\\
				GRU & 0.00412 & 0.00281 & 0.00315 & 0.00362\\
				\hline
			\end{tabular}
		\end{table}
		\subsection{Results}
		\begin{figure}[t]
			\centering
			\includegraphics[width = \linewidth, height=3cm]{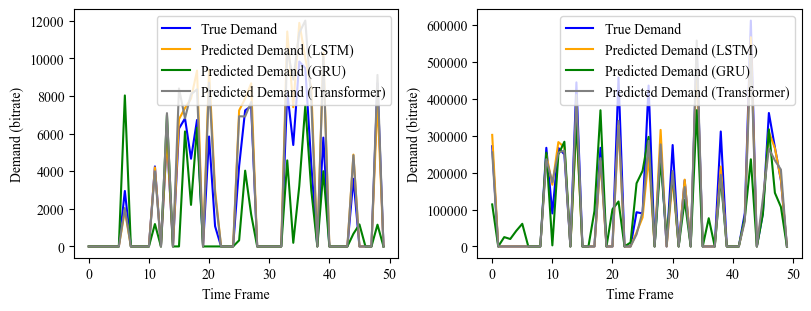}\\
			\centering { \hspace{0.5cm}(a) \hspace{4cm} (b) }\\
			\includegraphics[width = \linewidth, height=3cm]{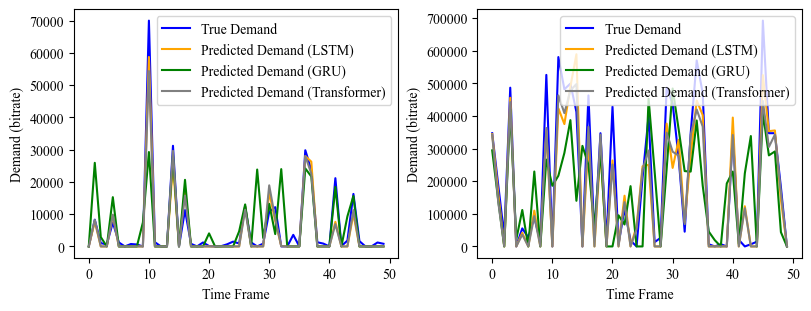}\\
			\centering {\hspace{0.5cm} (c)  \hspace{4 cm} (d) }\\
			\caption{Traffic demand forecasting results for two randomly selected user using our proposed Transformer-based model, compared against LSTM and GRU baselines. The plots illustrate the prediction performance for both eMBB and uRLLC traffic in an AI-RAN environment under static and mobile scenarios: (a) uRLLC demand for a static UE, (b) eMBB demand for a static UE, (c) uRLLC demand for a mobile UE, and (d) eMBB demand for a mobile UE.}
			
			\label{demand_ue}
		\end{figure}		
		We now present key results that demonstrate the effectiveness of the proposed framework. Table~\ref{demand_result} presents the MSE-based prediction errors for eMBB and uRLLC traffic demand, comparing our proposed Transformer method against baseline LSTM and GRU models. The Transformer model is unequivocally the top performer, achieving the lowest MSE in all four categories. This indicates its superior capability in capturing the temporal dependencies of both eMBB and uRLLC traffic patterns. Its most notable success is in predicting uRLLC traffic for moving UEs, where its MSE of 0.00288 is approximately 11\% lower than the next-best model, LSTM (0.00324). LSTM outperforms GRU in three of four cases. GRU performs well for static eMBB (MSE: 0.00311) but struggles with uRLLC, showing the highest errors (0.00492 static, 0.00542 moving), indicating its limitations in handling sporadic, critical traffic.
		
		Table~\ref{accuracy_result} reports service type identification accuracy for our proposed LSTM model comparing to GRU baseline and route selection accuracy for our proposed Transformer model comparing LSTM and GRU baseline. For route prediction, both LSTM and the Transformer are exceptionally accurate, with all reported scores above 99.8\%. The Transformer edges out LSTM in the moving UEs scenario (99.85\% vs. 99.81\%), a small but potentially significant margin in dynamic network management. For service prediction, LSTM demonstrates highest accuracy (99.86\% static, 99.81\% moving). The GRU model, however, shows a notable drop-off in performance. Its accuracy in the moving UEs scenario falls to 93.03\%, a significant decrease of over 6 percentage points compared to LSTM. This indicates a comparative weakness for GRU in correctly identifying services under dynamic network conditions.
		\begin{figure}[!t]
			\centering
			\includegraphics[ width= .45\linewidth]{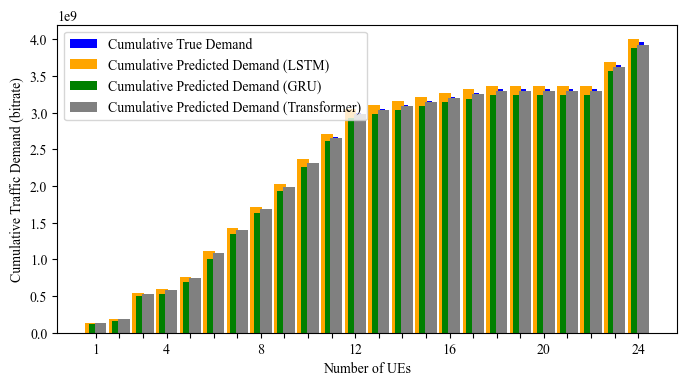}
			\includegraphics[ width= .45\linewidth]{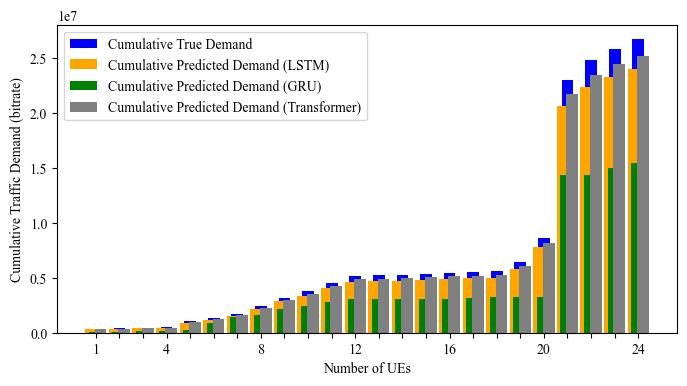}\\
			\centering { \hspace{0.5cm}(a) \hspace{4cm} (b)}\\
			\includegraphics[width= .45\linewidth]{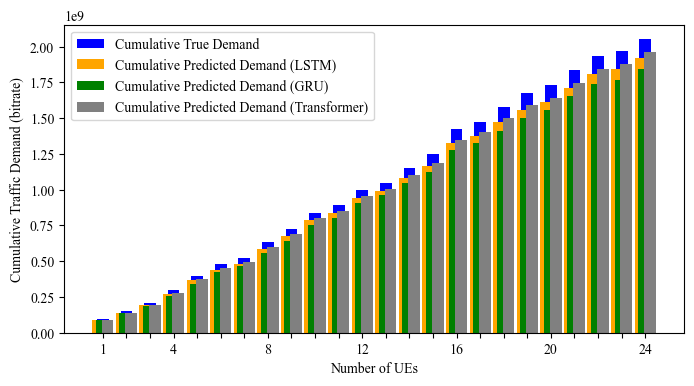}
			\includegraphics[ width= .45\linewidth]{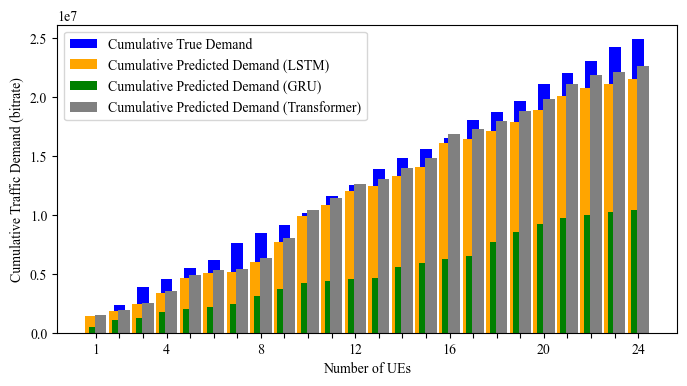}\\
			\centering {\hspace{0.5cm}(c) \hspace{4cm} (d)}\\
			\caption{Cumulative traffic demand prediction results with the increment of UEs for our proposed Transformer-based model, compared against LSTM and GRU baselines considering both static and mobile scenarios. The plots show cumulative prediction performance for eMBB and uRLLC traffic under different mobility conditions: (a) eMBB demand for static UE, (b) uRLLC demand for static UE, (c) eMBB demand for mobile UE, and (d) uRLLC demand for mobile UE.}
			\label{avg_demand}
		\end{figure}
		
		Table~\ref{resource_result} provides an MSE-based evaluation of the models' ability to predict the allocation of PRBs and power, comparing LSTM and GRU.	The findings here are definitive: LSTM consistently and significantly outperforms GRU in predicting resource allocation. For instance, in predicting PRBs for static UEs, LSTM's MSE of 0.00302 is over 26\% lower than GRU's MSE of 0.00412. This performance gap underscores LSTM's superior precision in forecasting fine-grained resource needs. Furthermore, LSTM demonstrates remarkable stability. Its prediction error for PRBs remains almost constant between the static (0.00302) and moving (0.00304) scenarios, suggesting its predictions are robust to UE mobility. GRU's performance, in contrast, is less stable and consistently less accurate.
		
		\begin{figure}[t]
			\centering
			\includegraphics[ width=\linewidth]{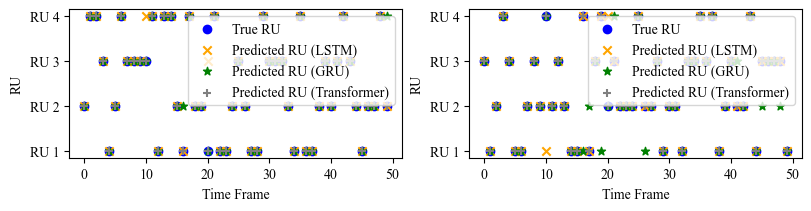}\\
			\centering {(a) \hspace{4cm} (b) }\\
			\caption{Traffic route prediction performance for a random UE using our proposed Transformer-based model, compared to LSTM and GRU baselines under both static and mobile scenarios. The plots illustrate route prediction under different mobility conditions: (a) Traffic routes for a static UE, and (b) Traffic routes for a mobile UE.}
			\label{ru_ue}
		\end{figure}
		\begin{figure}[!t]
			\centering
			\includegraphics[width= .48\linewidth]{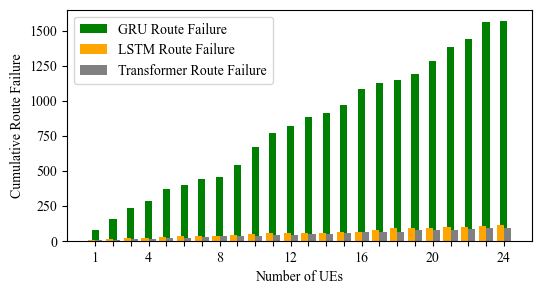}
			\includegraphics[width= .48\linewidth]{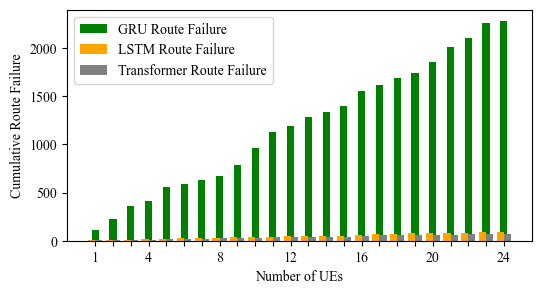}\\
			\centering {(a) \hspace{4cm} (b)}\\
			\caption{Traffic route failure cases of our proposed Transformer-based model compared to LSTM and GRU baselines, evaluated under both static and mobile scenarios as the number of UEs increases. The bars illustrate route failure cases under different mobility conditions: (a) Route failures for static UEs, and (b) Route failures for mobile UEs.}
			\label{avg_route_failure}
		\end{figure}
		Figs. \ref{demand_ue} and \ref{avg_demand} compare eMBB and uRLLC traffic demand predictions for static and mobile users for our proposed Transformer model comparing with the baseline LSTM, and GRU, against the ground truth. In Fig. \ref{demand_ue}, predictions over 50 time frames show that while traffic demand is highly unpredictable, the proposed Transformer method effectively forecasts traffic demand for the next time frame in both static and dynamic scenarios. Similarly, Fig. \ref{avg_demand} presents cumulative traffic demand predictions over 1646 time frames. For eMBB traffic, shown in charts (a) and (c), all three models demonstrate exceptional accuracy, as their predicted demand bars closely align with the true demand for both static and moving UEs. However, the performance differences become stark when predicting the more complex uRLLC traffic, as seen in charts (b) and (d). In these scenarios, the Transformer model stands out by delivering notable precise predictions that almost perfectly match the true demand, even under challenging mobile conditions. The LSTM model follows as a strong second, tracking the demand well but with a slight tendency to underestimate it. In stark contrast, the GRU model consistently and significantly underestimates the uRLLC demand. This critical flaw is most evident in the moving UE scenario (d), where the gap between the GRU’s prediction and the actual demand widens dramatically as the number of users increases. Ultimately, the figure provides powerful visual evidence that the Transformer is the most suitable and reliable architecture, while highlighting the GRU's significant limitations for predicting critical, low-latency traffic.
				
		Figs. \ref{ru_ue} and \ref{avg_route_failure} illustrate route predictions and corresponding failure rates for the next time frame under both static and moving scenarios. The results compare our proposed Transformer model with baseline LSTM and GRU models against the ground truth. Fig. \ref{ru_ue} shows route predictions over 50 time frames for randomly selected static and moving UEs. Despite the unpredictable fluctuations in route numbers, the proposed Transformer model consistently predicts appropriate routes for both scenarios. LSTM also performs well, ranking second to the Transformer, while GRU lags behind in prediction accuracy. Similarly, Fig. \ref{avg_route_failure} demonstrates the number of failures for route prediction or wrong route selection for static and dynamic scenarios. 
		\begin{figure}[t]
			\centering
			\includegraphics[width=\linewidth]{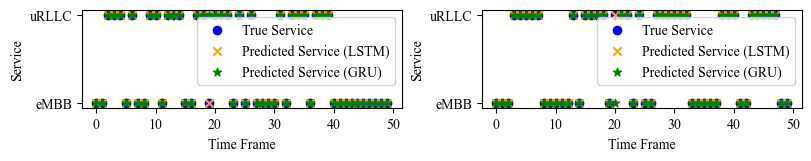}\\
			\centering {(a)  \hspace{4cm} (b) }\\
			\caption{Traffic service prediction results for a random UE over a mini time frame using our proposed LSTM model, compared to GRU baseline, evaluated under both static and mobile scenarios. The plots illustrate predicted traffic service types under different mobility conditions: (a) static UE and (b) mobile UE.}
			\label{service_ue}
		\end{figure}
		\begin{figure}[!t]
			\centering
			\includegraphics[width= .48\linewidth]{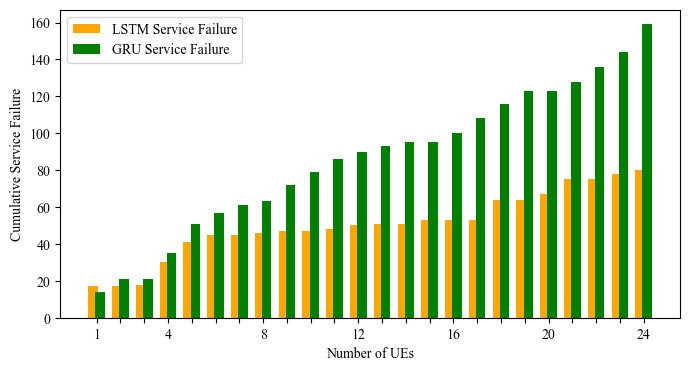}
			\includegraphics[width= .48\linewidth]{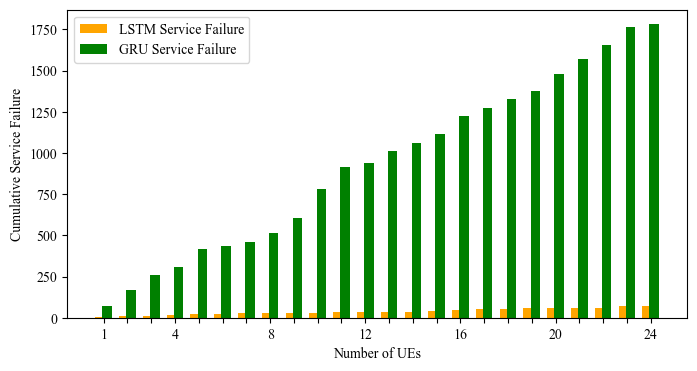}\\
			\centering {(a) \hspace{4cm} (b)}\\
			\caption{Service type prediction failure cases of our proposed LSTM model compared to GRU baseline, evaluated under both static and mobile scenarios as the number of UEs increases. The bars illustrate service identification failure cases under different mobility conditions: (a) Service type identification failure for static UE, and (b) Service type identification failure for mobile UE.}
			\label{avg_service_failure}
		\end{figure}
		
	    The key finding from the figure is the clear performance gap among models, with GRU as a distinct outlier. In both static and moving UE scenarios, GRU exhibits a sharp rise in route failures as the number of UEs increases, indicating poor scalability and reliability. In contrast, both LSTM and Transformer models maintain low cumulative failure rates, with the Transformer outperforming LSTM by incurring even fewer failures. This demonstrates that for route failure prediction, Transformer and LSTM are significantly more effective than GRU, with Transformer being the most reliable overall.
		\begin{figure}[t]
			\centering
			\includegraphics[width=\linewidth]{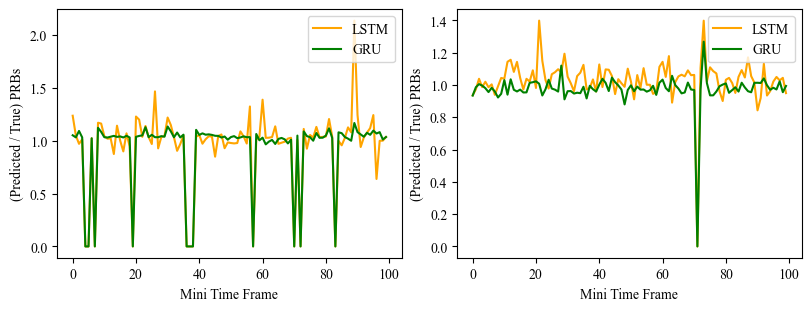}\\
			\centering {(a) \hspace{4cm} (b)}\\
			\caption{The ratio of predicted and true PRBs or a random user using our proposed LSTM model compared to GRU baseline considering both static and mobile scenarios. The graphs illustrate the ratio of predicted and true PRBs under different mobility conditions: (a) static UE, and (b) mobile UE.}
			\label{prb_ue}
		\end{figure}
		\begin{figure}[t]
			\centering
			\includegraphics[width= .48\linewidth]{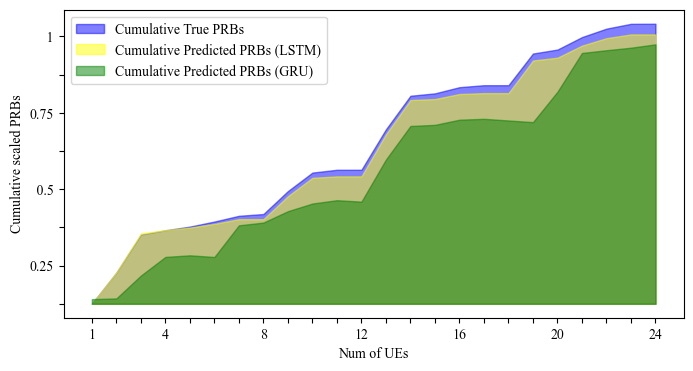}
			\includegraphics[width= .48\linewidth]{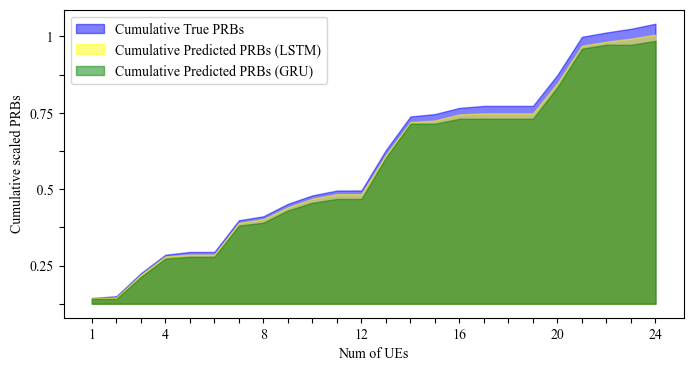}\\
			\centering {(a)  \hspace{4cm} (b) }\\
			\caption{Average PRBs assignment result using our proposed LSTM model compared to GRU baseline, evaluated under both static and mobile scenarios with the increment of UE. The graphs illustrate the predicted average PRBs under different mobility conditions: (a) static UE, and (b) mobile UE.}
			\label{avg_prb}
		\end{figure}
		
		Figs. \ref{service_ue} and \ref{avg_service_failure} present predictions of service types and failures for the next time frame under static and moving scenarios using proposed LSTM and baseline GRU models, compared to ground truth. In Fig. \ref{service_ue}, predictions over 50 time frames of one random static UE and one random moving UE show fluctuating service types in both stationary and dynamic user scenarios, with the proposed LSTM model accurately predicting service types despite high variability. Similarly, Fig. \ref{avg_service_failure} demonstrates the number of failures for service prediction or wrong service selection for static and dynamic scenarios. The figure highlights that the GRU model consistently exhibits a significantly higher cumulative service failure rate than the LSTM model in both static and moving UE scenarios. While service failures increase for both models as the number of UEs grows, GRU's failure rate remains substantially higher. This disparity becomes more pronounced in the moving UE scenario, where mobility introduces greater prediction challenges. LSTM shows greater robustness under these conditions.
		\begin{figure}[!t]
			\centering
			\includegraphics[width=\linewidth]{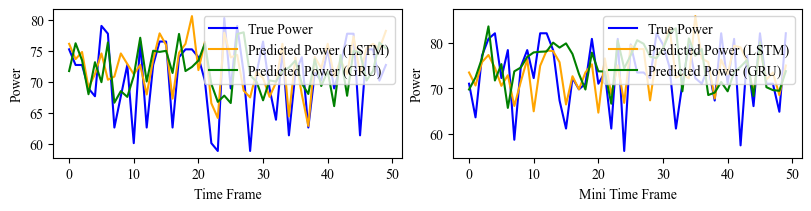}\\
			\centering {(a)  \hspace{4cm} (b) }\\
			\caption{Power prediction result using our proposed LSTM model compared to the baseline GRU model for a random user considering both static and mobile scenario. The graphs illustrate power prediction under different mobility conditions: (a) static UE, and (b) mobile UE.}
			\label{power_ue}
		\end{figure}
		
		\begin{figure}[t]
			\centering
			\includegraphics[width= .48\linewidth]{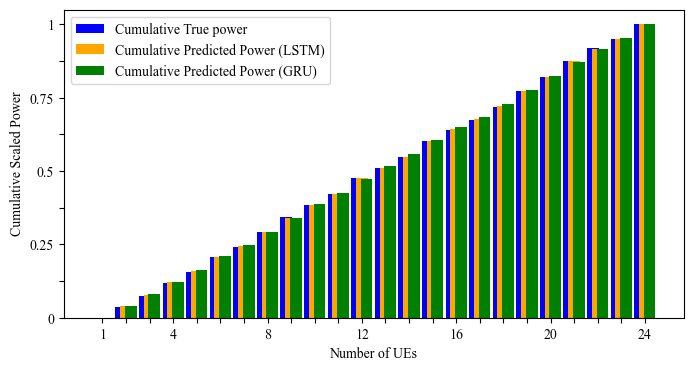}
			\includegraphics[width= .48\linewidth]{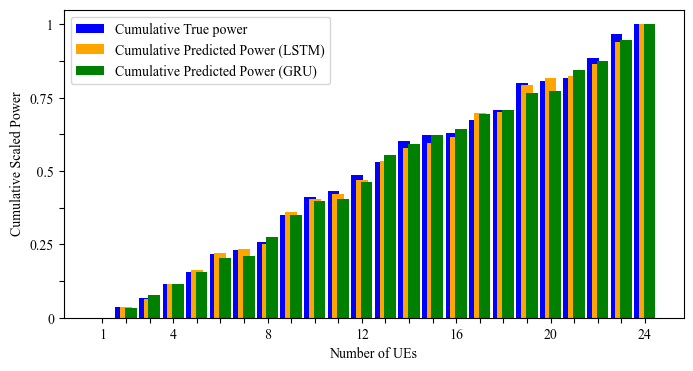}\\
			\centering {(a)  \hspace{4cm} (b) }\\
			\caption{Average power prediction result using our proposed LSTM model compared to GRU baseline, evaluated under both static and mobile scenarios with the increment of UE. The graphs illustrate the predicted average power under different mobility conditions: (a) static UE, and (b) mobile UE.}
			\label{avg_power}
		\end{figure}
		
		Figs. \ref{prb_ue} and \ref{avg_prb} depict PRB allocation predictions for static and dynamic scenarios using our proposed LSTM and baseline GRU models. In Fig. \ref{prb_ue}, predictions over 100 mini time frames show reliable performance despite fluctuating PRB demands. Fig. \ref{avg_prb} illustrates average predictions over 1512 mini time frames. In the static UE scenario (Fig. \ref{avg_prb}(a)), the LSTM model closely tracks the cumulative true PRB values, demonstrating accurate prediction as the number of UEs grows. In contrast, the GRU model consistently underestimates PRB assignments, with a noticeable gap from the ground truth that widens slightly as more UEs are added. This performance gap becomes more pronounced in the moving UE scenario (Fig. \ref{avg_prb}(b)), where the GRU model struggles significantly to match the true PRB values, highlighting its limitations in dynamic environments. Meanwhile, our proposed LSTM model maintains strong performance even under mobility, closely aligning with the ground truth
		
		\begin{figure}[t]
			\centering
			\includegraphics[width= .45\linewidth]{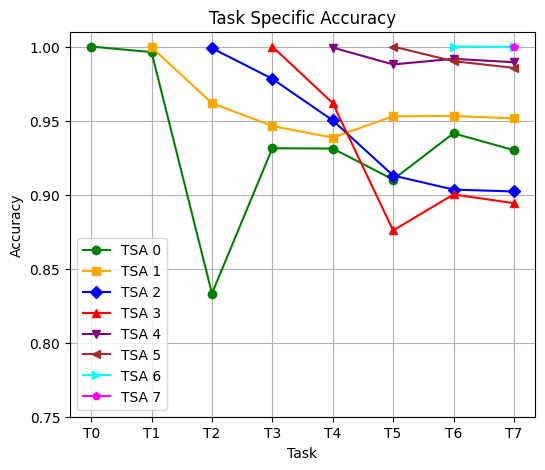}
			\includegraphics[width= .45\linewidth]{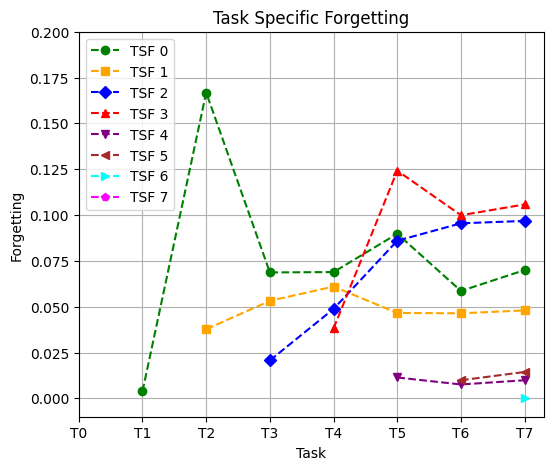}\\
			\text{(a)  \hspace{4cm} (b) }
			\caption{(a) Task specific accuracy, and (b) task specific forgetting for our proposed exemplar reply based continual learning method for continual service identification across eight sequential tasks.}
			\label{tsa}
		\end{figure}
		\begin{figure}[!t]
			\centering
			\includegraphics[width=.36\linewidth]{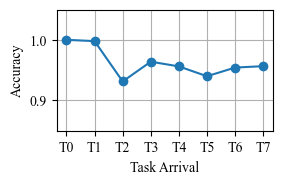}
			\includegraphics[width=.36\linewidth]{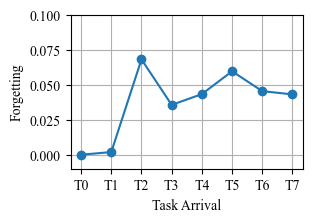}
			\text{(a)   \hspace{2cm} (b) }
			\caption{(a) Average accuracy, and (b) average forgetting for our proposed exemplar reply based continual learning method for continual service identification across eight sequential tasks.}
			\label{aaaf}
		\end{figure}
		
		Figs. \ref{power_ue} and \ref{avg_power} display the power predictions for one random static UE and one random moving UE for using proposed LSTM and baseline GRU models. In Fig. \ref{power_ue}, the models predict power consumption over 50 mini time frames, effectively handling fluctuating power levels. Fig. \ref{avg_power} shows average predictions for 1512 time frames. A key observation is that both LSTM and GRU models accurately predict power consumption, closely following the steady increase in cumulative true power as the number of UEs rises from 1 to 24. However, the proposed LSTM model aligns more closely with the true values than GRU, showing slightly higher precision. This high accuracy is consistently maintained even in the more complex moving UE scenario with dynamic network conditions.
	
		\begin{figure}[!t]
			\centering
			\includegraphics[width=.325\linewidth]{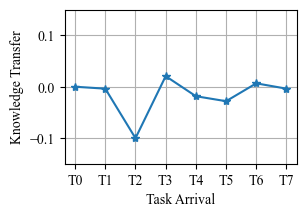}
			\includegraphics[width=.325 \linewidth]{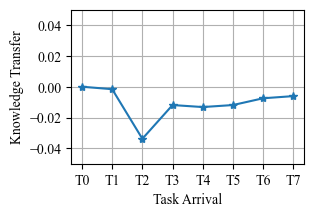}
			\includegraphics[width=.325\linewidth]{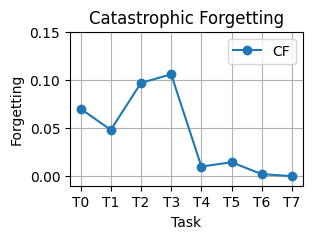}\\
			\text{(a) \hspace{2cm} (b)  \hspace{2cm} (c) }
			\caption{(a) Forward transfer, (b) backward transfer, and (c) catastrophic forgetting for our proposed exemplar reply based continual learning method for continual service identification across eight sequential tasks}
			\label{BWT}
		\end{figure}
		
		In Figs. \ref{tsa}, \ref{aaaf}, and \ref{BWT}, we present the continual learning performance of our proposed method. The results demonstrate a minimal decline in TSA and consistently low TSF across seven sequential tasks, which underscores the model’s strong ability to retain knowledge over time. As shown in Fig. \ref{tsa}, performance on previous tasks exhibits only slight degradation, while the model effectively recovers lost accuracy for tasks 0, 1, 2, and 3, successfully mitigating catastrophic forgetting. The trends in Fig. \ref{aaaf} further illustrate that while average accuracy and forgetting show modest declines, the model adapts dynamically and recovers performance in later tasks. Additionally, as depicted in Fig. \ref{BWT}, key continual learning metrics FWT, BWT, and CF confirm efficient knowledge transfer. Both FWT and BWT remain positive, while CF stays consistently low, demonstrating the model’s capability to retain past knowledge while seamlessly integrating new information.
		\begin{figure}[!t]
			\centering
			\includegraphics[ width=.7\linewidth, height=4cm]{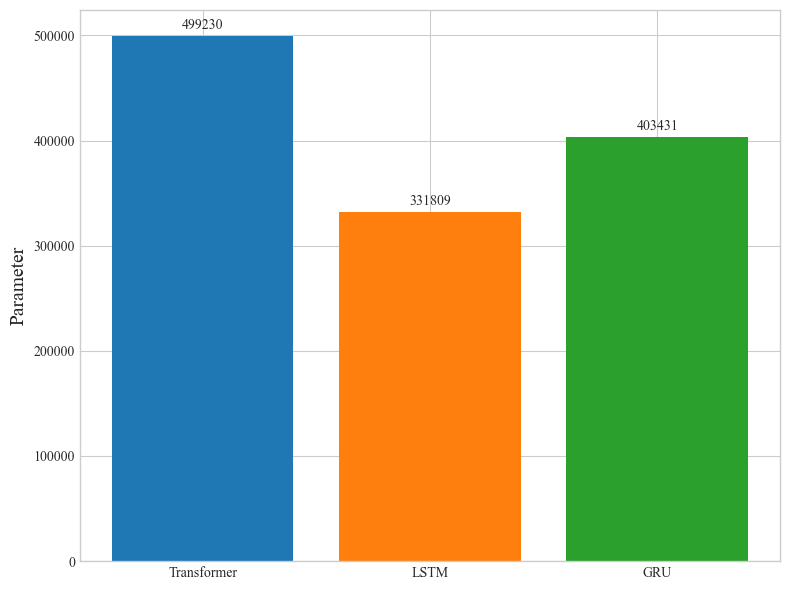}
			\caption{Comparison of number of parameter among Transformer, LSTM and GRU model.}
			\label{param}
		\end{figure}
		
		In Fig.~\ref{param}, we present the number of parameters for the Transformer, LSTM, and GRU models. The figure shows that the Transformer has 0.49M parameters, compared to 0.33M for LSTM and 0.40M for GRU. While the Transformer outperforms both LSTM and GRU in terms of prediction accuracy, it is slightly more resource-intensive. Due to this, we propose deploying the Transformer model as rApp1 within the non-RT RIC, where the execution times are longer than near RT RIC. In contrast, near-RT RIC environments require lightweight models capable of real-time operation. Therefore we propose LSTM there.  Also the performance of the xApp1 depends on the outputs of rApp1, we recommend using a ReVIN-empowered Transformer for rApp1 and a lightweight LSTM model for xApp1 to ensure efficient and responsive end-to-end processing across the RIC architecture.
		
		\section{Conclusion}\label{Con}
		We have proposed a novel deep incremental AI-RAN management framework designed to efficiently handle multi-service-modal UE (MSMU) supporting both eMBB and uRLLC services on a single UE. With the increasing demands of NextG networks, supporting multiple services on the same device is critical for applications such as the metaverse, where seamless high-speed throughput and low-latency communication are essential. Due to the non-linear and NP-hard nature of the formulated optimization problem for real-time RAN management, we decompose it into long-term and short-term subproblems and adopt an AI-driven solution. Specifically, we propose a ReVIN-enhanced Transformer model for traffic demand and route prediction, followed by heuristic-based resource slicing, and an incremental LSTM model for service identification and resource management. The experimental results validate that the proposed method effectively manages MSMU in the AI-RAN, achieving low mean square errors for traffic demand (0.0029), resource block prediction (0.003), and power prediction (0.002), while delivering 99\% accuracy in service type and route selection and over 95\% average accuracy for continual service adaptation across seven tasks. These findings underscore the practicality of our approach in addressing the challenges of NextG networks.
		

		
		\bibliographystyle{IEEEtran}
		\bibliography{arXiv.bib}

		\vfill
		
	\end{document}